\documentclass[prd,aps,nofootinbib,notitlepage,showpacs,preprintnumbers]{revtex4-1}
\usepackage{graphicx,epsf,amsmath,amsfonts,amssymb,amsbsy}
\usepackage{epsfig}
\newcommand{\ds}{\displaystyle}
\newcommand{\vev}[1]{\langle#1\rangle}
\newcommand{\mat}{\left ( \begin{array}}
\newcommand{\emat}{\end{array} \right )}
\newcommand{\vect}{\left ( \begin{array}{c}}
\newcommand{\evect}{\end{array} \right )}

\begin{document}

\title{Interplay between superconductivity and chiral symmetry breaking in
a (2+1)-dimensional model with compactified spatial coordinate}

\author{D. Ebert $^{1)}$,
T.G. Khunjua $^{2)}$, K.G. Klimenko $^{3)}$, and V.Ch. Zhukovsky $^{2)}$}
\vspace{1cm}

\affiliation{$^{1)}$ Institute of Physics,
Humboldt-University Berlin, 12489 Berlin, Germany}
\affiliation{$^{2)}$ Faculty of Physics, Moscow State University,
119991, Moscow, Russia} \affiliation{$^{3)}$ State Research Center
of Russian Federation -- Institute for High Energy Physics,
NRC "Kurchatov Institute", 142281, Protvino,
Moscow Region, Russia} 
\begin{abstract}
In this paper a (2+1)-dimensional model with four-fermion
interactions is investigated in the case when one spatial coordinate
is compactified and the space topology takes the form of an infinite
cylinder, $R^1\otimes S^1$. It is supposed that the system is
embedded into real three-dimensional space and that a magnetic flux
$\Phi$ crosses the transverse section of the cylinder. The model
includes four-fermion interactions both in the fermion-antifermion
(or chiral) and fermion-fermion (or superconducting) channels. We
then study phase transitions in dependence on the chemical potential
$\mu$ and the flux $\Phi$ in the leading order of the large-$N$
expansion technique, where $N$ is the number of fermion fields. It
is demonstrated that for arbitrary relations between coupling
constants in the chiral and superconducting channels,
superconductivity   appears in the system at rather high values of
$\mu$ (the length $L$ of the circumference $S^1$ is fixed).
Moreover, it is shown that at sufficiently small values of $\mu$ the
growth of the magnetic flux $\Phi$ leads to a periodical reentrance
of the chiral symmetry breaking or superconducting phase, depending
on the values of $\mu$ and coupling constants.
\end{abstract}
\maketitle

\section{Introduction}

It is well known that quantum field theories with four-fermion
interactions (4FQFT) play an essential role in several branches of
modern physics. In the case of (3+1)-dimensional QCD, effective
theories of this type are used in order to describe the low energy
physics of light mesons \cite{volkov} as well as phase transitions
in compact stars and in hadronic matter under the influence of
various external conditions such as temperature, magnetic fields
etc. (see, e.g., the review papers
\cite{buballa,ferrer,gatto,andrianov,hiller,andersen}). Low
dimensional 4FQFTs also find important applications in condensed
matter physics. For example, (1+1)-dimensional 4FQFTs, known as
Gross-Neveu models \cite{gross}, are suitable for the description of
polyacetylene-like systems \cite{caldas1}. In addition, due to their
renormalizability, asymptotic freedom and spontaneous breaking of
chiral symmetry, Gross-Neveu type models can be used as a laboratory
for the qualitative simulation of specific properties of QCD. In
particular, such effects of dense baryonic matter as color
superconductivity \cite{chodos,abreu,ekkz}, charged pion
condensation \cite{gubina1,gubina} and dynamical chiral symmetry
breaking \cite{thies,thies2} were investigated in the simplified
case of (1+1)-dimensional Gross-Neveu models.

Nowadays of special interest are (2+1)-dimensional 4FQFTs. These
models mimic main properties of corresponding (3+1)-dimensional
models. Thus, in the framework of (2+1)-dimensional models, one has investigated corresponding phenomena as dynamical symmetry breaking
\cite{gross,semenoff,rosenstein,klimenko2,inagaki,hands}, color
superconductivity \cite{toki}, and QCD-motivated phase diagrams \cite{kneur}. Other examples of this kind are spontaneous chiral symmetry breaking induced by external magnetic/chromomagnetic fields (this effect was for the first time studied  also in terms of (2+1)-dimensional 4FQFT \cite{klimenko}) as well as gravitational catalysis of chiral symmetry breaking \cite{gies}. It is worth mentioning that these theories are
also useful in developing new QFT techniques like e.g. the optimized perturbation theory \cite{kneur,k}.

However, there is yet another and more physical motivation for
studying (2+1)-dimensional 4FQFTs. It is based on the fact that many
condensed matter systems have a (quasi-)planar structure. Among
these systems are the high-T$_c$ cuprate and iron superconductors
\cite{davydov}, and the one-atom thick layer of carbon atoms, or
graphene, \cite{niemi,castroneto}. Thus, many properties of these
planar physical systems can be explained on the basis of various
(2+1)-dimensional models, including the 4FQFTs (see, e.g.,
\cite{Semenoff:1998bk,caldas,roy,zkke,marino,kzz,Cao:2014uva} and references therein). In particular, the influence of such external factors as temperature, chemical potential and magnetic field on the
metal-to-insulator phase transition and quantum Hall effect in
planar fermionic systems is investigated in the framework of 4FQFTs
(see, e.g., in \cite{caldas,roy}). Another example is
superconductivity of planar condensed matter systems which can also
be treated qualitatively in terms of (2+1)-dimensional 4FQFTs
\cite{marino,kzz}.

The (2+1)-dimensional 4FQFT model of the papers \cite{marino,kzz}
describes a competition between two processes: chiral symmetry
breaking (excitonic pairing) and superconductivity (Cooper pairing).
Its structure is a direct generalization of the known
(1+1)-dimensional 4FQFT model of Chodos et al. \cite{chodos,abreu},
which remarkably mimics the temperature $T$ and chemical potential
$\mu$ phase diagram of real QCD, to the case of (2+1)-dimensional
spacetime. Recall that in \cite{chodos,abreu}, in order to avoid the
prohibition on Cooper pairing as well as spontaneous breaking of a
continuous $U(1)$ symmetry in (1+1)-dimensional models (known as the
Mermin-Wagner-Coleman no-go theorem \cite{coleman}), the
consideration was performed in the leading order of the
$1/N$-technique, i.e. in the large-$N$ limit assumption, where $N$
is the number of fermion fields. In this case, quantum fluctuations,
which would otherwise destroy a long-range order corresponding to
spontaneous symmetry breaking, are suppressed by $1/N$ factors. By
the same reason in the (2+1)-dimensional 4FQFT model of the papers
\cite{marino,kzz} and in the case of finite values of $N$,
spontaneous breaking of continuous $U(1)$ symmetry is allowed only
at zero temperature, i.e. it is forbidden at $T>0$. One possible way
to enable the investigation of superconducting phase transitions in
the framework of this (2+1)-dimensional model at $T>0$ is to use the
constraint $N\to\infty$, as it was done in \cite{chodos,abreu}.

The present paper is devoted to the investigation of the competition
between excitonic and Cooper pairing of fermionic quasiparticles in
the framework of the above mentioned (2+1)-dimensional 4FQFT model
under influence of a chemical potential $\mu$ \cite{marino,kzz}. In
contrast to these papers, where a flat two-dimensional space with
trivial topology $R^2$ was used, we now suppose that the space
topology is nontrivial and has the form $R^1\otimes S^1$, where the
length of the circumference $S^1$ is denoted by $L$. Thus, in our
consideration one spatial coordinate is compactified. (Note that
(1+1)- and (3+1)-dimensional 4FQFT models of superconductivity
with compactified spatial coordinates were already studied in
\cite{abreu,ebert}.) We hope that the investigation of a rather
special four-fermionic system on a cylindrical surface will be
useful for the understanding of physical processes taking place,
e.g., in carbon nanotubes.

The paper is organized as follows. In Sec. II the (2+1)-dimensional
4FQFT model, which describes interactions in the fermion-antifermion
(or chiral) and fermion-fermion (or superconducting) channels is
presented. Here the unrenormalized thermodynamic potential (TDP) of
the model is obtained in the leading order of the large-$N$
expansion technique (see the Subsec. II A). In Subsec. II B a
renormalization group invariant expression for the TDP is obtained
whose global minimum point provides us with chiral and Cooper pair
condensates. The phase portrait of the model is presented in Fig. 1
in the case $L=\infty$, $\mu=0$. In Sec. III a renormalization group
invariant expression for the TDP is obtained in the case $L\ne
\infty$. Here the system is considered as an infinite cylinder,
embedded into real three-dimensional space. In addition, it is
supposed that there is a magnetic flux $\Phi$ through the transverse
section of the cylinder. In the next Sec. IV typical phase diagrams
Fig.2 and Fig. 3 of the model at $L\ne\infty$ and $\mu=0$ are
presented at $0\le\phi<1/6$ and $1/6 <\phi<1/2$, correspondingly,
where $\phi=\Phi/\Phi_0$ ($\Phi_0$ is the elementary magnetic field
flux). Here a duality between chiral symmetry breaking and
superconductivity phenomena is observed. Moreover, it is shown that,
depending on the relation between coupling constants, a periodical
reentrance of chiral symmetry breaking or superconducting phases (as
well as periodic symmetry restoration) occurs with growing values of
the magnetic flux $\Phi$. At $L\ne \infty$ and $\mu\ne 0$ the phase
structure of the model is investigated in Sec. V. It is established
here that if there is an arbitrary small attractive interaction in
the fermion-fermion channel, then it turns out possible to generate
the superconductivity phenomenon in the system by increasing the
chemical potential. Some related technical problems of our
consideration are relegated to three Appendices.

\section{The case $L\to\infty$}
\subsection{ The model and its thermodynamic potential}
\label{effaction}

Our investigation is based on a (2+1)-dimensional 4FQFT model
with massless fermions belonging to a fundamental multiplet of the
auxiliary  $O(N)$ flavor group. Its Lagrangian describes the
interaction both in the scalar fermion--antifermion (or chiral) and scalar
difermion (or superconducting) channels \cite{chodos}:
\begin{eqnarray}
 {\cal L}=\sum_{k=1}^{N}\bar \psi_k\Big [\gamma^\nu i\partial_\nu
+\mu\gamma^0\Big ]\psi_k&+& \frac {G_1}N\left (\sum_{k=1}^{N}\bar
\psi_k\psi_k\right )^2+\frac {G_2}N\left (\sum_{k=1}^{N}
\psi_k^TC\psi_k\right )\left (\sum_{j=1}^{N}\bar
\psi_jC\bar\psi_j^T\right ), \label{1}
\end{eqnarray}
where $\mu$ is the fermion number chemical potential. As noted
above, all fermion fields $\psi_k$ ($k=1,...,N$) form a fundamental
multiplet of the $O(N)$ group. Moreover, each field $\psi_k$ is a
four-component Dirac spinor (the symbol $T$ denotes the
transposition operation). The quantities $\gamma^\nu$ ($\nu =0,1,2$)
and $\gamma^5$ are matrices in the 4-dimensional spinor space.
Moreover, $C\equiv\gamma^2$ is the charge conjugation matrix. The
algebra of the $\gamma$-matrices as well as their particular
representation are given in Appendix \ref{ApA}. Clearly, the
Lagrangian ${\cal L}$ is invariant under transformations from the
internal $O(N)$ group, which is introduced here in order to make it
possible to perform all the calculations in the framework of the
nonperturbative large-$N$ expansion method. Physically more
interesting is that the model (1) is invariant under $U(1)$ group
transformations demonstrating  fermion number conservation
$\psi_k\to\exp (i\alpha)\psi_k$ ($k=1,...,N$), and that there is a
symmetry of the model under discrete $\gamma^5$ chiral
transformation:  $\psi_k\to\gamma^5\psi_k$ ($k=1,...,N$).

The 'linearized', i.e. with only quadratic powers of fermionic fields,
version of Lagrangian (\ref{1}) that
contains auxiliary scalar bosonic fields $\sigma (x)$, $\pi (x)$, $\Delta(x)$,
$\Delta^{*}(x)$ has the following form
\begin{eqnarray}
\widetilde {\cal L}\ds&=&\bar\psi_k\Big [\gamma^\nu i\partial_\nu
+\mu\gamma^0 -\sigma \Big ]\psi_k-\frac{N\sigma^2}{4G_1}-\frac N{4G_2}\Delta^{*}\Delta-
 \frac{\Delta^{*}}{2}[\psi_k^TC\psi_k]
-\frac{\Delta}{2}[\bar\psi_k C\bar\psi_k^T].
\label{3}
\end{eqnarray}
(Here and in what follows summation over repeated indices
$k=1,...,N$ is implied.) Clearly, the Lagrangians (\ref{1}) and
(\ref{3}) are equivalent, as can be seen by using the Euler-Lagrange
equations of motion for scalar bosonic fields which take the form
\begin{eqnarray}
\sigma (x)=-2\frac {G_1}N(\bar\psi_k\psi_k),~~
\Delta(x)=-2\frac {G_2}N(\psi_k^TC\psi_k),~~
\Delta^{*}(x)=-2\frac {G_2}N(\bar\psi_k C\bar\psi_k^T).
\label{4}
\end{eqnarray}
One can easily see from (\ref{4}) that the (neutral) field
$\sigma(x)$ is a real quantity, i.e.
$(\sigma(x))^\dagger=\sigma(x)$ (the
superscript symbol $\dagger$ denotes the Hermitian conjugation), but
the (charged) difermion scalar fields $\Delta(x)$ and $\Delta^*(x)$
are Hermitian conjugated complex quantities, so
$(\Delta(x))^\dagger= \Delta^{*}(x)$ and vice versa. Clearly, all
the fields (\ref{4}) are singlets with respect to the $O(N)$ group.
\footnote{Note that the $\Delta (x)$ field is a flavor O(N) singlet,
since the representations of this group are real.} If the scalar
difermion field $\Delta(x)$ has a nonzero ground state expectation
value, i.e.\  $\vev{\Delta(x)}\ne 0$, then the abelian fermion number
$U(1)$ symmetry of the model is spontaneously broken down and
superconductivity (SC) appears in the system. However,
if $\vev{\sigma (x)}\ne 0$ then chiral symmetry breaking (CSB) phase
is realized spontaneously in the model.

We begin our investigation of the phase structure of the
four-fermion model (1) using the equivalent semi-bosonized
Lagrangian (\ref{3}). In the leading order of the large-$N$ (mean
field) approximation, the effective action ${\cal S}_{\rm
{eff}}(\sigma,\pi,\Delta,\Delta^{*})$ of the model under
consideration is expressed by means of the path integral over
fermion fields:
$$
\exp(i {\cal S}_{\rm {eff}}(\sigma,\Delta,\Delta^{*}))=
  \int\prod_{l=1}^{N}[d\bar\psi_l][d\psi_l]\exp\Bigl(i\int \widetilde {\cal
  L}\,d^3 x\Bigr),
$$
where
\begin{eqnarray}
&&{\cal S}_{\rm {eff}} (\sigma,\Delta,\Delta^{*}) =-\int
d^3x\left [\frac{N}{4G_1}\sigma^2(x)+
\frac{N}{4G_2}\Delta (x)\Delta^{*}(x)\right ]+ \widetilde {\cal
S}_{\rm {eff}}. \label{5}
\end{eqnarray}
The fermion contribution to the effective action, i.e.\  the term
$\widetilde {\cal S}_{\rm {eff}}$ in (\ref{5}), is given by
\begin{equation}
\exp(i\widetilde {\cal S}_{\rm
{eff}})=\int\prod_{l=1}^{N}[d\bar\psi_l][d\psi_l]\exp\Bigl\{i\int\Big
[\bar\psi_k(\gamma^\nu i\partial_\nu +\mu\gamma^0-\sigma)\psi_k -
 \frac{\Delta^{*}}{2}(\psi_k^TC\psi_k)
-\frac{\Delta}{2}(\bar\psi_k C\bar\psi_k^T)\Big ]d^3
x\Bigr\}. \label{6}
\end{equation}
The ground state expectation values $\vev{\sigma(x)}$,
$\vev{\Delta(x)}$, and $\vev{\Delta^*(x)}$ of the composite bosonic
fields are determined by the saddle point equations,
\begin{eqnarray}
\frac{\delta {\cal S}_{\rm {eff}}}{\delta\sigma (x)}=0,~~~~~
\frac{\delta {\cal S}_{\rm {eff}}}{\delta\Delta (x)}=0,~~~~~
\frac{\delta {\cal S}_{\rm {eff}}}{\delta\Delta^* (x)}=0. \label{7}
\end{eqnarray}
From now on we suppose that the quantities $\vev{\sigma(x)}$,
$\vev{\Delta(x)}$, and $\vev{\Delta^*(x)}$ do not depend on space
coordinates, i.e. $\vev{\sigma(x)}=M$, $\vev{\Delta(x)}=\Delta$ and
$\vev{\Delta^*(x)}=\Delta^*$, where $M,\Delta,\Delta^*$ are constant
quantities. In fact, the quantities $M$, $\Delta$ and $\Delta^*$ are
coordinates of the global minimum point of the thermodynamic
potential (TDP) $\Omega (M,\Delta,\Delta^*)$. In the leading order
of the large $N$-expansion it is defined by the following
expression:
\begin{equation*}
\int d^3x\Omega (M,\Delta,\Delta^*) =-\frac{1}{N}{\cal S}_{\rm
{eff}}\{\sigma(x),\Delta (x),\Delta^*(x)\}\Big|_{\sigma (x)=M,\Delta (x)=\Delta,\Delta^*(x)=\Delta^*} ,
\end{equation*}
which gives
\begin{eqnarray}
\int d^3x\Omega (M,\Delta,\Delta^*)\,\,&=&\,\,\int d^3x\left
(\frac{M^2}{4G_1}+\frac{\Delta\Delta^*}{4G_2}\right )+\frac{i}{N}\ln\left
( \int\prod_{l=1}^{N}[d\bar\psi_l][d\psi_l]\exp\Big (i\int d^3 x\Big
[\bar\psi_k D\psi_k\right.\nonumber\\&& \left.-
\frac{\Delta}{2}(\psi_k^TC\psi_k)
-\frac{\Delta^*}{2}(\bar\psi_k C\bar\psi_k^T)\Big ]
\Big )\right ), \label{9}
\end{eqnarray}
where $D=\gamma^\rho i\partial_\rho +\mu\gamma^0-M$. To proceed, let
us first point out that without loss of generality the quantities
$\Delta,\Delta^*$ might be considered as real ones.
\footnote{Otherwise,  phases of the complex quantities
$\Delta,\Delta^*$ might be eliminated by an appropriate
transformation of fermion fields in the path integral (\ref{9}).}
So, in the following we will suppose that
$\Delta=\Delta^*\equiv\Delta$, where $\Delta$ is already a real
quantity. Then, in order to find a convenient expression for the TDP
it is necessary to invoke Appendix B of \cite{kzz}, where a path
integral similar to (\ref{9}) is evaluated. Taking into account this
technique, we obtain the following expression for the zero
temperature, $T=0$, TDP of the 4FQFT model (1):
\begin{eqnarray}
\Omega (M,\Delta)=
\frac{M^2}{4G_1}+\frac{\Delta^2}{4G_2}
+i\int\frac{d^3p}{(2\pi)^3}\ln\Big [(p_0^2-(E^+)^2)(p_0^2
-(E^-)^2)\Big ], \label{12}
\end{eqnarray}
where the notation $\Omega (M,\Delta)$ is now used for the TDP
$\Omega (M,\Delta,\Delta^*)$ at $\Delta=\Delta^*\equiv\Delta$,
$(E^\pm)^2=E^2+\mu^2+\Delta^2\pm 2 \sqrt{M^2\Delta^2+\mu^2E^2}$ and
$E=\sqrt{M^2+p_1^2+p_2^2}$. Obviously, the function $\Omega
(M,\Delta)$ is invariant under each of the transformations $M\to-M$,
$\Delta\to -\Delta$ and $\mu\to-\mu$.  Hence, without loss of
generality, it is sufficient to restrict ourselves by the
constraints $M\ge 0$, $\Delta\ge 0$ and $\mu\ge 0$ and to
investigate the properties of the TDP (\ref{12}) just in this
region. Using a rather general formula
\begin{eqnarray}
\int_{-\infty}^\infty dp_0\ln\big
(p_0-A)=\mathrm{i}\pi|A|,\label{int}
\end{eqnarray}
(see e.g., Appendix B of \cite{gubina};
the relation(9) is true up to an infinite term independent of the
real quantity $A$), it is possible to reduce the expression
(\ref{12}) to the following one:
\begin{eqnarray}
\Omega (M,\Delta)\equiv\Omega^{un} (M,\Delta)=
\frac{M^2}{4G_1}+\frac{\Delta^2}{4G_2}
-\int\frac{d^2p}{(2\pi)^2}\Big (E^++E^-\Big ).  \label{13}
\end{eqnarray}
Note that the following asymptotic expansion is valid:
\begin{eqnarray}
E^++E^-=2|\vec p|+\frac{M^2+\Delta^2}{|\vec p|}+{\cal O}(1/|\vec
p|^3),\label{16}
\end{eqnarray}
where $|\vec p|=\sqrt{p_1^2+p_2^2}$. Hence the integral term in
(\ref{13}) is ultraviolet divergent, and $\Omega (M,\Delta)$ is an
unrenormalized quantity. Hence, in (\ref{13}) and below we  use the
equivalent notation $\Omega^{un} (M,\Delta)$ for it.

\subsection{Renormalization procedure and phase structure at $\mu=0$}
\label{mu0}

To renormalize the TDP (\ref{13}) it is useful to rewrite this quantity
in the following way
\begin{eqnarray}
\Omega^{un} (M,\Delta)&=&
\frac{M^2}{4G_1}+\frac{\Delta^2}{4G_2}-\int\frac{d^2p}{(2\pi)^2}\left
(E^+\big |_{\mu=0}+E^-\big |_{\mu=0}\right )
-\int\frac{d^2p}{(2\pi)^2}\Big (E^++E^--E^+\big |_{\mu=0}-E^-\big
|_{\mu=0}\Big ), \label{013}
\end{eqnarray}
where
 \begin{eqnarray*}
E^+\big |_{\mu=0}+E^-\big |_{\mu=0}=
\sqrt{|\vec p|^2+(M+\Delta)^2}+\sqrt{|\vec p|^2+(M-\Delta)^2}.
\end{eqnarray*}
Since the leading terms of the asymptotic expansion (\ref{16}) do not
depend on $\mu$, it is clear that the last integral in (\ref{013}) is
a convergent one. Other terms in (\ref{013}) form the unrenormalized
TDP $V^{un}(M,\Delta)$ (effective potential) at $\mu=0$,
\begin{eqnarray}
V^{un} (M,\Delta)&=&
\frac{M^2}{4G_1}+\frac{\Delta^2}{4G_2}-\int\frac{d^2p}{(2\pi)^2}\left
( \sqrt{|\vec p|^2+(M+\Delta)^2}+\sqrt{|\vec
p|^2+(M-\Delta)^2}\right ) ,\label{14}
\end{eqnarray}
i.e. the expression (\ref{013}) has the following equivalent form
 \begin{eqnarray}
\Omega^{un} (M,\Delta)=V^{un}(M,\Delta)
-\int\frac{d^2p}{(2\pi)^2}\Big (E^++E^-- \sqrt{|\vec
p|^2+(M+\Delta)^2}-\sqrt{|\vec p|^2+(M-\Delta)^2}\Big ).\label{15}
\end{eqnarray}
Thus, to renormalize the TDP (\ref{13})--(\ref{15}) it is sufficient
to remove the ultraviolet divergency from the effective potential
$V^{un}(M,\Delta)$ (\ref{14}). This procedure is performed, as e.g.,
in \cite{kzz} and based on the special $\Lambda$ dependence of the
bare coupling constants $G_1$ and $G_2$ (here $\Lambda$ is the
cutoff parameter of the integration region in (\ref{14}),
$|p_1|<\Lambda$ and $|p_2|<\Lambda$),
\begin{eqnarray}
\frac 1{4G_1}\equiv \frac
1{4G_1(\Lambda)}=\frac{2\Lambda\ln(1+\sqrt{2})}{\pi^2}+\frac{1}{2\pi
g_1}, ~~~\frac 1{4G_2}\equiv \frac
1{4G_2(\Lambda)}=\frac{2\Lambda\ln(1+\sqrt{2})}{\pi^2}+\frac{1}{2\pi
g_2}, \label{018}
\end{eqnarray}
where $g_{1,2}$ are finite and $\Lambda$-independent model
parameters with dimensionality of the inverse mass. Moreover, since
bare couplings $G_1$ and $G_2$ do not depend on a normalization
point, the same property is also valid for $g_{1,2}$. As a result,
upon cutting the integration region in (\ref{14}) and using there
the substitution (\ref{018}), it becomes possible to obtain in the
limit $\Lambda\to\infty$ the following renormalized expression
$V^{ren}(M,\Delta)$ for the effective potential of the model in
vacuum (for more details see \cite{kzz}),
\begin{eqnarray}
V^{ren}(M,\Delta)\equiv \Omega^{ren}(M,\Delta)\big |_{\mu=0}=
\frac{M^2}{2\pi g_1}+\frac{\Delta^2}{2\pi
g_2}+\frac{(M+\Delta)^{3}}{6\pi}+\frac{|M-\Delta|^{3}}{6\pi}.\label{25}
\end{eqnarray}
It should also be mentioned that the TDP (\ref{25}) is a
renormalization group invariant quantity. Investigating the behavior
of the global minimum point (GMP) of the TDP (\ref{25}) with the
coupling constants $g_1$ and $g_2$, it is possible to establish the
corresponding phase portrait of the model (1) at $L=\infty$ and
$\mu=0$ \cite{kzz} (see Fig. 1). In this figure the notations I, II,
and III mean the symmetric, the CSB, and the SC phase,
correspondingly. In the symmetric phase the GMP of the TDP
(\ref{25}) lies at the point $(M=0,\Delta=0)$, and the initial
symmetry of the model (1) remains intact. In the phase II the GMP is
of the form $(M=-1/g_1,\Delta=0)$, which means spontaneous breaking
of the $\gamma^5$ chiral symmetry in the ground state of the system.
Finally, in the superconducting phase III the GMP looks like
$(M=0,\Delta=-1/g_2)$. As a result, the $U(1)$ symmetry of the model is spontaneously broken down.
\begin{figure}
\includegraphics[width=0.45\textwidth]{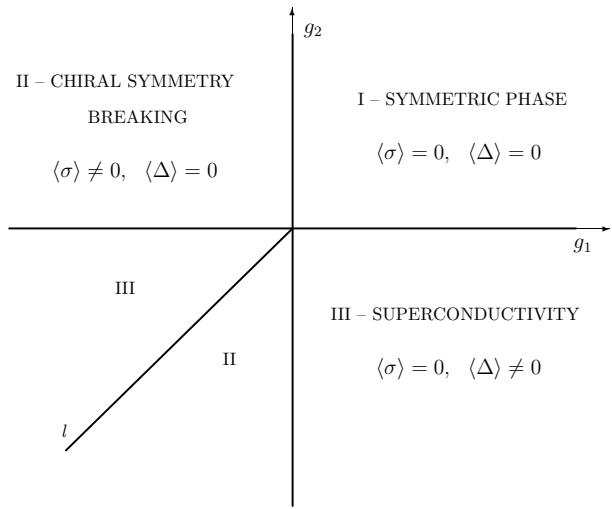}
\caption{The $(g_1,g_2)$-phase portrait of the model at $\mu=0$ and
$L=\infty$. The notations I, II and III mean the symmetric, the
chiral symmetry breaking (CSB) and the superconducting (SC) phases,
respectively. At $g_{1,2}<0$ the line {\it l} is defined by the
relation {\it l} $\equiv\{(g_1,g_2):g_1=g_2\}$. }
\end{figure}
Clearly, if the cutoff parameter $\Lambda$ is fixed, then the phase
structure of the model can be described in terms of bare coupling
constants $G_1, G_2$ instead of finite quantities $g_1, g_2$.
Indeed, let us first introduce a critical value of the couplings,
$G_c=\frac{\pi^2}{8\Lambda\ln(1+\sqrt{2})}$. Then, as it follows
from Fig. 1 and (\ref{018}), at $G_1<G_c$ and $G_2<G_c$ the symmetric
phase I of the model occurs. If $G_1>G_c$, $G_2<G_c$ ($G_1<G_c$,
$G_2>G_c$), the CSB phase II (the SC phase III) is
realized. Finally, let us suppose that both $G_1>G_c$ and
$G_2>G_c$. In this case at $G_1>G_2$ ($G_1<G_2$) we have again the CSB
phase II (the SC phase III).

The fact that it is possible to renormalize the effective potential
of the initial model (1) in the leading order of the large
$N$ expansion is the reflection of a more general property of
(2+1)-dimensional theories with four-fermion interactions. Indeed,
it is well known that in the framework of 'naive'
perturbation theory in coupling constants these models are not
renormalizable. However, as it was proved in \cite{rosenstein}
in the framework of the nonperturbative large-$N$ technique, 
these models are renormalizable in each order of the $1/N$ expansion.

Now, it is evident that a renormalized expression $\Omega^{ren}(M,\Delta)$ for the TDP of the model looks like
 \begin{eqnarray}
\Omega^{ren} (M,\Delta)=V^{ren}(M,\Delta)
-\int\frac{d^2p}{(2\pi)^2}\Big (E^++E^-- \sqrt{|\vec
p|^2+(M+\Delta)^2}-\sqrt{|\vec p|^2+(M-\Delta)^2}\Big ). \label{015}
\end{eqnarray}
Just this quantity should be used in order to establish the phase structure of the model at $L=\infty$ and $\mu\ne 0$.
\label{IIB}

\section{TDP in the case $L\ne\infty$}

In the present Section we start the investigation of the
difermion and fermion-antifermion condensations in the framework of
(2+1)-dimensional 4FQFT model (1), when one of two spatial coordinates
is compactified. \footnote{Note that in \cite{8,gamayun,kolmakov,kim}
a phase structure of a more simple special case of the
(2+1)-dimensional model (1), i.e. at $G_2=0$, was investigated in
spaces with different nontrivial topologies. The impact of finite
size effects, curvature of space etc. on the chiral symmetry
breaking was considered also in \cite{flachi} in spaces of
different dimensions on the basis of the zeta-function
regularization method.} In this case, the two-dimensional space has a
nontrivial topology of the form $R^1\otimes S^1$. Without loss of
generality, it is supposed here that just the $y$-axis is compactified and fermion fields satisfy boundary conditions of the form (the $x$ coordinate
is not restricted)
\begin{eqnarray}
\psi_k(t,x,y+L)=e^{i2\pi\phi}\psi_k(t,x,y), \label{016}
\end{eqnarray}
where $L$ is the length of the circumference $S^1$ and $k=1,...,N$.

The physical situation in our consideration can be treated in the
following way. We suppose that in  real three-dimensional space there
is a two-dimensional surface on which the physical system, described
by Lagrangian (1), is located. The surface is then rolled into an
infinite cylinder $R^1\otimes S^1$ and, in addition, a homogeneous external magnetic field $B$ parallel to the cylinder axis is switched on. So the magnetic flux $\Phi=\pi r^2B$ pervades through the transverse
section of the cylinder (here $r$ is the radius of the circumference
$S^1$, $r=L/(2\pi)$). In this case one can imagine that the magnetic
phase $\phi$ in (\ref{016}) is the quantity $\phi=\Phi/\Phi_0$, where
$\Phi_0=2\pi/e$ is the elementary magnetic field flux. Just this
interpretation of the quantity $\phi$ is taken e.g., in \cite{gamayun} and 
in the present paper. 
\footnote{\label{foot}
In real physical systems the boundary conditions (\ref{016}) might
slightly change. For example, for carbon nanotubes the phase in the boundary conditions (\ref{016}) is changed, $\phi\to\alpha+\phi$, where
$\alpha =0$ for metallic nanotubes and $\alpha =\pm 1/3$ for
semiconducting ones \cite{ando} (here $\phi$ is still the quantity
$\phi=\Phi/\Phi_0$). } 
Do not be confused, but for shortness we use in the following the same name ``magnetic flux`` both for the genuine magnetic flux $\Phi$ and for the ratio $\phi=\Phi / \Phi_0$.

In this case, to obtain the (unrenormalized) thermodynamic potential
$\Omega^{un}_{L\phi}(M,\Delta)$ of the initial system, one must simply 
replace the integration over the $p_2$-momentum in
(\ref{013})-(\ref{15}) by an infinite series, using the rule
\begin{eqnarray}
\int_{-\infty}^{\infty}\frac{dp_2}{2\pi}f(p_{2})\to\frac
1L\sum_{n=-\infty}^{\infty}f(p_{n\phi}),~~~~p_{n\phi}=
\frac{2\pi}{L}(n+\phi),~~~n=0,\pm 1, \pm 2,... \label{17}
\end{eqnarray}
So we have from (\ref{15})
 \begin{eqnarray}
\Omega^{un}_{L\phi}(M,\Delta)&=&V^{un}_{L\phi}(M,\Delta)
-\frac 1L\int\frac{dp_1}{2\pi}\sum_{n=-\infty}^{\infty}\Bigg (
E_{nL\phi}^++E_{nL\phi}^-\nonumber\\
&-&\sqrt{p_1^2+\frac{4\pi^2(n+\phi)^2}{L^2}+(M+\Delta)^2}-\sqrt{p_1^2+
\frac{4\pi^2(n+\phi)^2}{L^2}+(M-\Delta)^2}~\Bigg ), \label{18}
\end{eqnarray}
where
\begin{eqnarray}
E_{nL\phi}^\pm=\sqrt{p_1^2+\frac{4\pi^2(n+\phi)^2}{L^2}+M^2+\mu^2+\Delta^2\pm
2 \sqrt{M^2\Delta^2+\mu^2\left
(p_1^2+\frac{4\pi^2(n+\phi)^2}{L^2}+M^2\right )}}\label{19}
\end{eqnarray}
and $V^{un}_{L\phi}(M,\Delta)$ is analogously obtained from 
expression (\ref{14}) by using the transformation rule (\ref{17}),
\begin{eqnarray}
V^{un}_{L\phi} (M,\Delta)= \frac{M^2}{4G_1}+\frac{\Delta^2}{4G_2}&-&
\frac 1L\int\frac{dp_1}{2\pi}\sum_{n=-\infty}^{\infty}\Bigg (
\sqrt{p_1^2+\frac{4\pi^2(n+\phi)^2}{L^2}+(M+\Delta)^2}\nonumber\\
&+&\sqrt{p_1^2+\frac{4\pi^2(n+\phi)^2}{L^2}+(M-\Delta)^2}~\Bigg ).
\label{20}
\end{eqnarray}
It is clear from (\ref{18})-(\ref{20}) that the TDP is a periodic
function with unit period with respect to the magnetic flux $\phi$. So
in many respects it is enough to study the TDP (\ref{18}) only at
$-1/2\le\phi\le 1/2$. We have proved in Appendix \ref{ApB} that the
expression (\ref{20}) is equal to the following one,
\begin{eqnarray}
V^{un}_{L\phi} (M,\Delta)=V^{un}(M,\Delta) +\frac 1{\pi
L^3}\sum_{\pm}\sum_{n=1}^{\infty}\frac{\exp
(-nL|M\pm\Delta|)}{n^3}\Big (nL|M\pm\Delta|+1\Big )\cos(2\pi n\phi),
\label{B8}
\end{eqnarray}
where $V^{un}(M,\Delta)$ is the unrenormalized  effective potential
of the model in vacuum, i.e. at $\mu=0$ and $L=\infty$ (see the
expression (\ref{14}) or, alternatively, (\ref{C3})). Since the
remaining integral and series terms both in (\ref{18}) and (\ref{B8}) are
convergent, it is clear that in order to obtain finite renormalized
expression $\Omega^{ren}_{L\phi}(M,\Delta)$ for the TDP at $L\ne\infty$, 
one should simply remove the ultraviolet divergency from
the vacuum effective potential $V^{un}(M,\Delta)$, using the way of
Sec. \ref{IIB}. As a result we have from (\ref{18}) and (\ref{B8})
 \begin{eqnarray}
&&\Omega^{ren}_{L\phi}(M,\Delta)=V^{ren}(M,\Delta)
+\frac 1{\pi L^3}\sum_{\pm}\sum_{n=1}^{\infty}
\frac{\exp (-nL|M\pm\Delta|)}{n^3}\Big (nL|M\pm\Delta|+1\Big )
\cos(2\pi n\phi)~~~~~~~21\nonumber\\
&-&\frac 1L\int\frac{dp_1}{2\pi}\sum_{n=-\infty}^{\infty}\Bigg
(E_{nL\phi}^++E_{nL\phi}^-
-\sqrt{p_1^2+\frac{4\pi^2(n+\phi)^2}{L^2}+(M+\Delta)^2}-
\sqrt{p_1^2+\frac{4\pi^2(n+\phi)^2}{L^2}+(M-\Delta)^2}~\Bigg ),
\label{21}
\end{eqnarray}
where $V^{ren}(M,\Delta)$ is given in (\ref{25}). Moreover, one can
see from (\ref{21}) that in addition to the constraints $M\ge 0$,
$\Delta\ge 0$ and $\mu\ge 0$ (see the comments just after
(\ref{12})) it is enough to accept, without loss of generality, 
the restriction $0\le\phi\le 1/2$ as well. \footnote{This restriction is
a consequence of the symmetry of the TDP (\ref{21}) with respect to
the transformation $\phi\to -\phi$.}

\section{Phase structure at $L\ne\infty$ and $\mu=0$}

In this case the TDP (\ref{21}) has a simpler form,
 \begin{eqnarray}
V^{ren}_{L\phi}(M,\Delta)&\equiv&\Omega^{ren}_{L\phi}(M,\Delta)\big
|_{\mu=0}\nonumber\\&=&V^{ren}(M,\Delta) +\frac 1{\pi
L^3}\sum_{\pm}\sum_{n=1}^{\infty}\frac{\exp
(-nL|M\pm\Delta|)}{n^3}\Big (nL|M\pm\Delta|+1\Big )\cos(2\pi n\phi).
\label{021}
\end{eqnarray}
Numerical investigations show that a global minimum  point (GMP) of
the TDP (\ref{021}) cannot be located at the point of the form
$(M\ne 0,\Delta\ne 0)$, i.e. at least one of the quantities $M$ and
$\Delta$ is equal to zero in the GMP of the effective potential. So,
in order to establish the GMP $(M_0,\Delta_0)$ of the effective
potential (\ref{021}), it is sufficient to compare the least values
of the simpler functions, $F_{1\phi}(M)$ and $F_{2\phi}(\Delta)$,
which are the reductions of the effective potential
$V^{ren}_{L\phi}(M,\Delta)$ on the $M$ and $\Delta$ axis,
correspondingly. Evidently,
\begin{eqnarray}
F_{1\phi}(M)\equiv V^{ren}_{L\phi}(M,\Delta=0) &=&
\frac{M^2}{2\pi g_1}+\frac{M^{3}}{3\pi}+
\frac 2{\pi L^3}\sum_{n=1}^{\infty}\frac{\exp (-nLM)}{n^3}\Big (nLM+1\Big )\cos(2\pi n\phi),\label{022}\\
F_{2\phi}(\Delta)\equiv
V^{ren}_{L\phi}(M=0,\Delta)&=&\frac{\Delta^2}{2\pi
g_2}+\frac{\Delta^{3}}{3\pi}+\frac 2{\pi
L^3}\sum_{n=1}^{\infty}\frac{\exp (-nL\Delta)}{n^3}\Big
(nL\Delta+1\Big )\cos(2\pi n\phi).\label{023}
\end{eqnarray}
Let us find the GMP $M_0$ of the function $F_{1\phi}(M)$ as well as
its properties depending on the external parameters $L$, $\phi$ and
$g_1$. For this we need the stationary, or gap, equation,
\begin{eqnarray}
\label{0180} \frac{\partial F_{1\phi}(M)}{\partial M}=0=\frac M\pi
f(M)\equiv \frac M\pi\left\{\frac 1{g_1} +M +\frac 1{L} \ln\Big [1
+e^{-2LM}-2e^{-LM}\cos(2\pi\phi)\Big ]\right\}.
\end{eqnarray}
It is easy to see that $f(M)$ from (\ref{0180}) is a monotonically
increasing function at $M\ge 0$. Moreover, $f(\infty)=\infty$.
Hence, apart from a trivial solution $M=0$, there exists a single
nonzero solution $M_0\ne 0$ of the equation (\ref{0180}) if and only
if $f(0)<0$, i.e. at
\begin{eqnarray}
\label{019} \frac 1{g_1} +\frac 2L\ln [2\sin (\pi\phi)] <0.
\end{eqnarray}
It is evident that just under the condition (\ref{019}) the point
$M_0\ne 0$ is a global minimum point of the function $F_{1\phi}(M)$.
Solving the equation (\ref{0180}), one can find in this case that
\begin{eqnarray}
M_0(L)=\frac 1L\hbox{arccosh}\left
(\cos(2\pi\phi)+\frac{e^{-L/g_1}}2\right ),\label{020}
\end{eqnarray}
where $\hbox{arccosh}(x)=\ln(x+\sqrt{x^2-1})$ is the function
defined at $x\ge 1$. If the condition (\ref{019}) is not satisfied,
then the stationary equation (\ref{0180}) has only a trivial solution
$M=0$, which is the GMP of the effective potential (\ref{022}) in
this case.

Similar properties are valid for the function (\ref{023}). Namely,
if
\begin{eqnarray} \label{0190} \frac 1{g_2} +\frac 2L\ln [2\sin
(\pi\phi)] <0,
\end{eqnarray}
then its GMP is at the nonzero point
\begin{eqnarray}
\Delta_0(L)=\frac 1L\hbox{arccosh}\left
(\cos(2\pi\phi)+\frac{e^{-L/g_2}}2\right ).\label{0200}
\end{eqnarray}
However, if the constraint (\ref{0190}) is violated, then we have a least value of the function (\ref{023}) at the trivial point $\Delta=0$.

Now, comparing the minima of the functions (\ref{022}) and
(\ref{023}), it is possible to find both the genuine GMP of TDP
(\ref{021}) and its dependence on the external parameters. As a
result, one can establish the phase structure of the model. By this
way, we have obtained the $(g_1,g_2)$-phase diagrams of the model at
arbitrary fixed values of $L\ne \infty$ and magnetic flux $\phi$
(see Figs 2 and 3). In the phases I, II and III of these figures the
GMP of the TDP (\ref{021}) has the form $(0,0)$, $(M_0(L),0)$ and
$(0,\Delta(L))$, respectively (the gaps $M_0(L)$ and
$\Delta_0(L)$ are defined by the relations (\ref{020}) and
(\ref{0200}), respectively). So in phase I the initial symmetries of
the model remain intact, in phase II the chiral symmetry is
broken down, whereas in phase III there is superconductivity in the ground state. Due to the periodicity property of the model with
respect to $\phi$, our investigations are restricted to
the region $0\le\phi\le 1/2$. Moreover, it turns out that for different values of $\phi$ from this region we have quite different phase diagrams. Indeed, Fig. 2 presents the $(g_1,g_2)$-phase structure at $0\le\phi< 1/6$, whereas the phase diagram in Fig. 3 corresponds to magnetic flux
values from the region $1/6<\phi\le 1/2$. The quantity $g_c$ in
these figures is the solution of the equation $f(0)=0$ with respect
to the coupling constant $g_1$ (the function $f(x)$ is defined in
(\ref{0180})), 
\begin{eqnarray} 
\label{190} g_c=-\frac{L}{2\ln [2\sin
(\pi\phi)]}.
\end{eqnarray}
On the lines $g_1=g_c$ or $g_2=g_c$ there are second order phase
transitions from chiral symmetry breaking II or superconducting
phase III to the symmetrical phase I. In contrast, the line $l$ of
these figures corresponds to a first order phase transitions between
II and III phases. Moreover, it is clear from (\ref{190}) that at 
$\phi\to 1/6_\pm$ we have $g_c\to\mp\infty$,
i.e. $g_c$ is not a finite quantity. So the phase diagram in the
case $\phi=1/6$ can not be represented by Figs 2 and 3. Note
that in this particular case the $(g_1,g_2)$-phase structure of the
model looks formally like in Fig. 1, where $L=\infty$. However, it
is evident that the order parameters, or gaps $\vev{\sigma}$ and
$\vev{\Delta}$, corresponding to these particular cases of the
phase structure of the model are quite different. Indeed, in the
case of Fig. 1 with $L=\infty$ we have $\vev{\sigma}=-1/g_1$ and
$\vev{\Delta}=-1/g_2$, whereas in the case $L\ne\infty$ and
$\phi=1/6$ the gaps are presented by the relations (\ref{020}) or
(\ref{0200}).

It follows from (\ref{190}) that the critical coupling constant
$g_c$ varies in the interval $0<g_c<\infty$ when $0<\phi<1/6$.
However, at $1/6<\phi\le 1/2$ we have the following constraint on
the critical value $g_c$: $-\infty<g_c\le g_0\equiv -L/[2\ln 2]$.
Taking into account these observations, it is possible to construct
with the help of Figs. 2 and 3 the evolution of the phase structure of
the model with respect to a magnetic flux $\phi$ at arbitrary fixed
values of $L\ne\infty$ and coupling constants $g_1$ and $g_2$.
Indeed, if the point $(g_1,g_2)$ belongs to  the strips $g_0 \equiv
-L/[2\ln 2]<g_1<0$
and/or $g_0<g_2<0$, then, as it is clear from Figs. 2 and 3, we have
CSB or SC phases for all values of $\phi$. Moreover, in each point of
these strips the phase structure of the model is not changed vs
$\phi$. It means that in this
case order parameters (\ref{020}) or (\ref{0200}) are positively
defined and periodic functions vs $\phi$ (see Fig. 4, where the
graphic of the order parameter $M_0(L)$ vs $\phi$ is presented for
$g_1=-L/\ln 6$, i.e. at $g_0<g_1<0$). 

However, the situation is different for points from another regions of the $(g_1,g_2)$ plane, i.e. when a point $(g_1,g_2)$ belongs to one of the following regions, (i) $\{(g_1,g_2):g_1>0,g_2>0\}$, (ii)
$\{(g_1,g_2):g_1>0,g_2<g_0\}$, (iii)
$\{(g_1,g_2):g_1<g_0,g_2<g_0\}$, and (iv)
$\{(g_1,g_2):g_1<g_0,g_2>0\}$. Indeed, in this case at $\phi=0$ the
initial symmetry is spontaneously broken down at any finite values
of L (see the phase portrait of Fig. 2 with $g_c=0$). Then, with
growing value of $\phi$ the gap of the CSB or SC phase
decreases and at some critical value $\phi_c$, where $0<\phi_c<1/2$,
becomes zero. At this moment there is a restoration by a 2nd order
phase transition of the initial symmetry of the model. Note that for
all values of the magnetic flux $\phi$ such that $\phi_c<\phi<1-\phi_c$ the system is in its symmetric phase I. After that at $\phi=1-\phi_c$
there again appears a phase with broken symmetry, and a gap
increases in the interval $1-\phi_c<\phi<1$. In the following the
process is periodically repeated. In Fig. 5 the behavior of the gap
$\Delta_0(L)$ vs $\phi$ is shown, when $(g_1,g_2)$ belongs to the
above mentioned region (i), where in addition we suppose that
$g_2=2L$ and $g_1<g_2$ (in this case at $\phi=0$ the SC phase is
realized in the model). For such choice of the coupling constants
we have $\phi_c\approx 0.13$.

Hence, we see that if the coupling constants $g_1$ and $g_2$ are fixed
inside one of the strips $g_0<g_1<0$ and/or $g_0<g_2<0$, where $g_0=
-L/[2\ln 2]$, then for all values of the magnetic flux $\phi$ the
symmetry of the ground state of the model persists to be the same as
at $\phi =0$ (i.e. phase structure of the model does not change vs
$\phi$).  However,
in this case there is an oscillation of the gap vs $\phi$ (see Fig.
4 for an illustration). For the rest points of the $(g_1,g_2)$ plane,
the increasing of the external magnetic flux $\phi$ along the
axis of a cylinder is accompanied by periodical reentering of CSB or
SC phase (which one depends on the point $(g_1,g_2)$) as well as with
periodical reentering of a symmetry restoration. Such effects, if
exist, can be observed experimentally.
\begin{figure}
\includegraphics[width=0.45\textwidth]{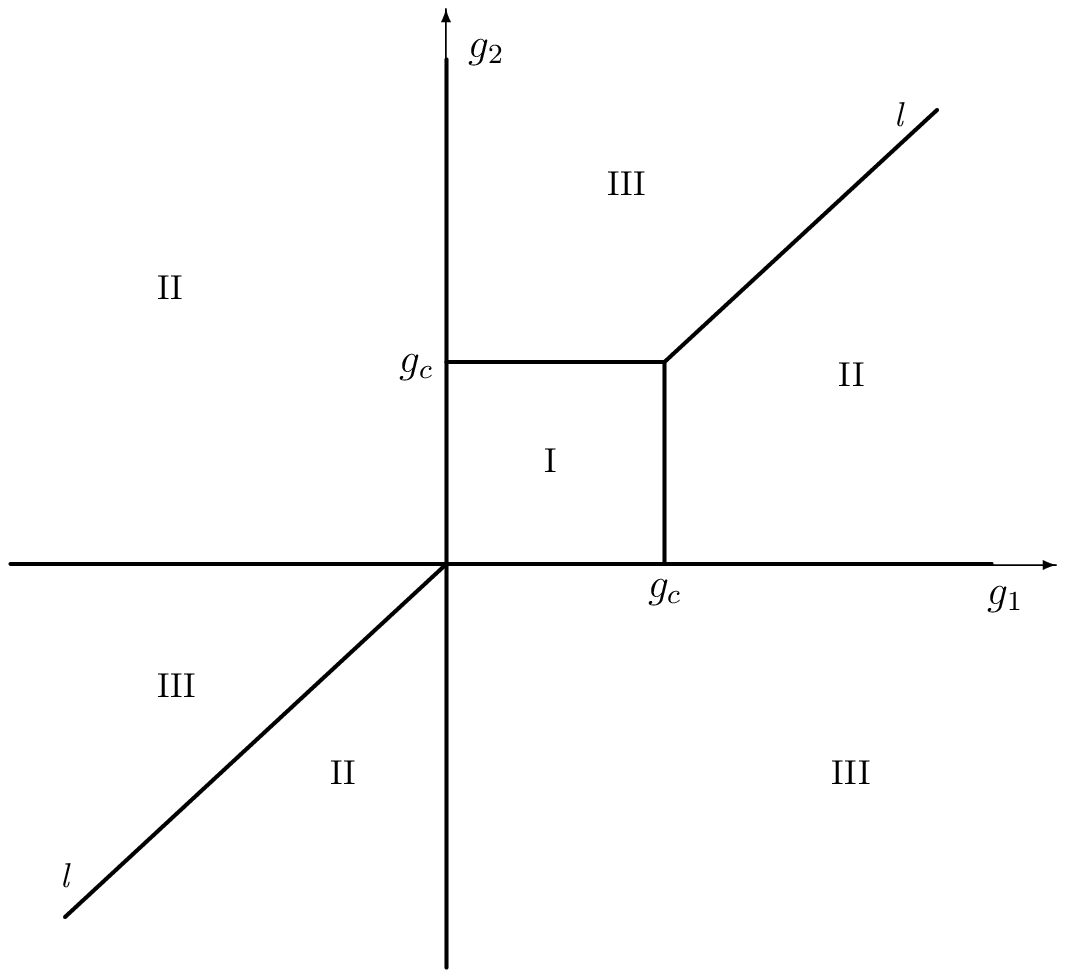}
\hfill
\includegraphics[width=0.43\textwidth]{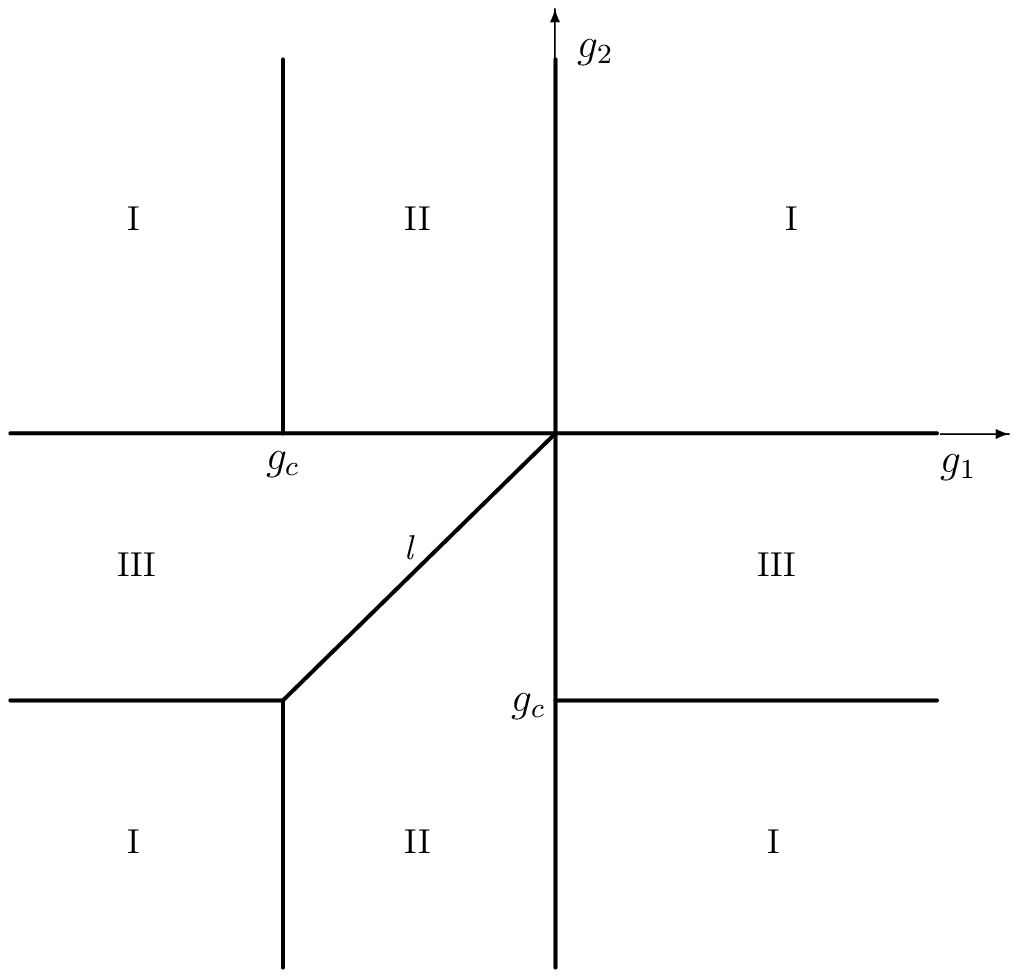}\\
\parbox[t]{0.45\textwidth}{
\caption{The $(g_1,g_2)$-phase portrait of the model at $\mu=0$ and
fixed values of $L\ne\infty$ and $\phi$, where $0\le\phi<1/6$. We
use the same designations of the phases as in Fig. 1. In the regions
$g_{1,2}<0$ and $g_{1,2}>g_c$, where $g_c$ is presented in
(\ref{190}), the line {\it l}  is defined by the relation  {\it l}
$\equiv\{(g_1,g_2):g_1=g_2\}$. }}\hfill
\parbox[t]{0.45\textwidth}{
\caption{The $(g_1,g_2)$-phase portrait of the model at $\mu=0$ and
fixed values of $L\ne\infty$ and $\phi$, where $1/6<\phi<1/2$. We
use the same designations of the phases as in Fig. 1. The line {\it
l}  is defined by the relation  {\it l}
$\equiv\{(g_1,g_2):g_1=g_2\}$. Critical value $g_c$ is presented in
(\ref{190}).} }
\end{figure}
\begin{figure}
\includegraphics[width=0.45\textwidth]{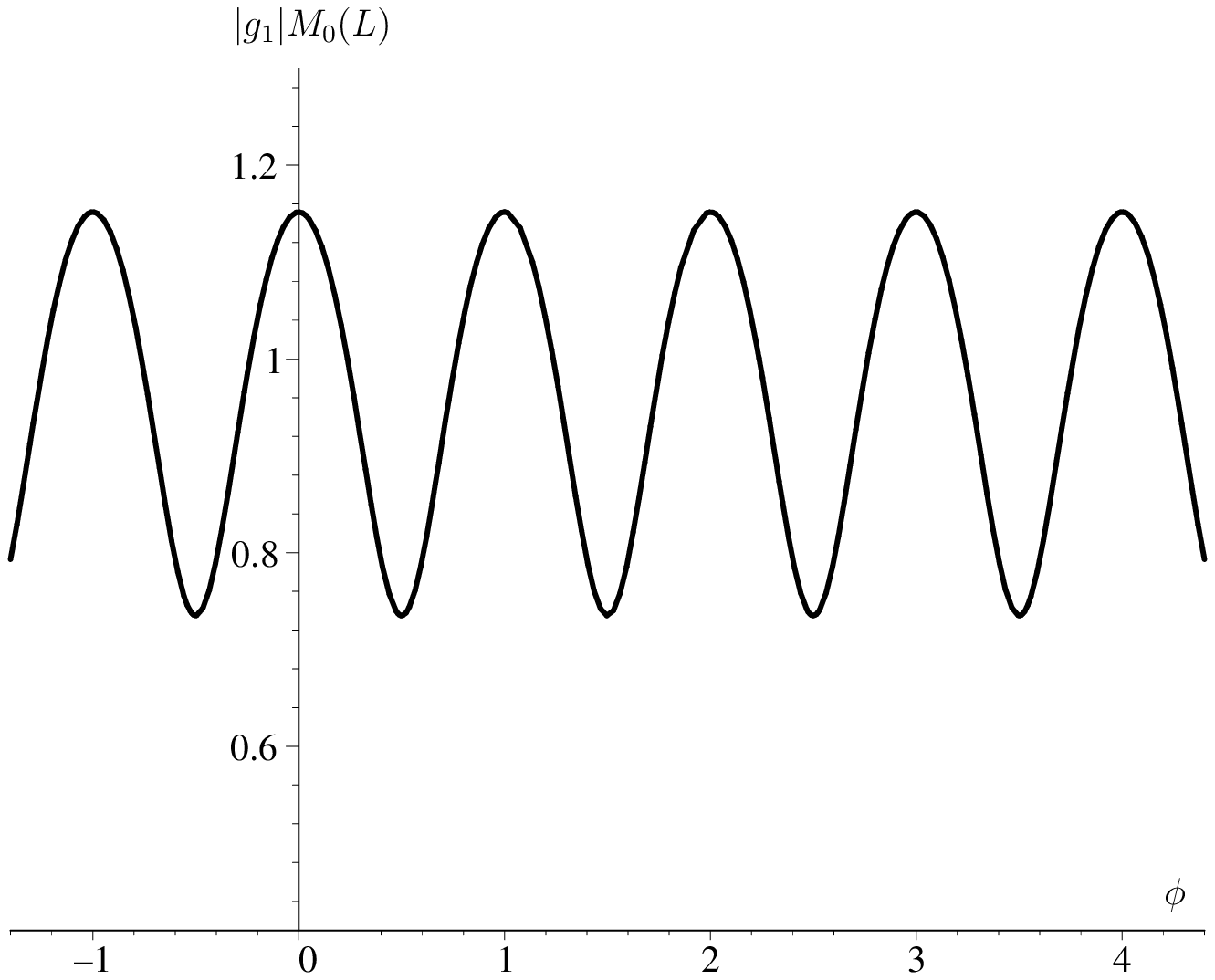}
\hfill
\includegraphics[width=0.45\textwidth]{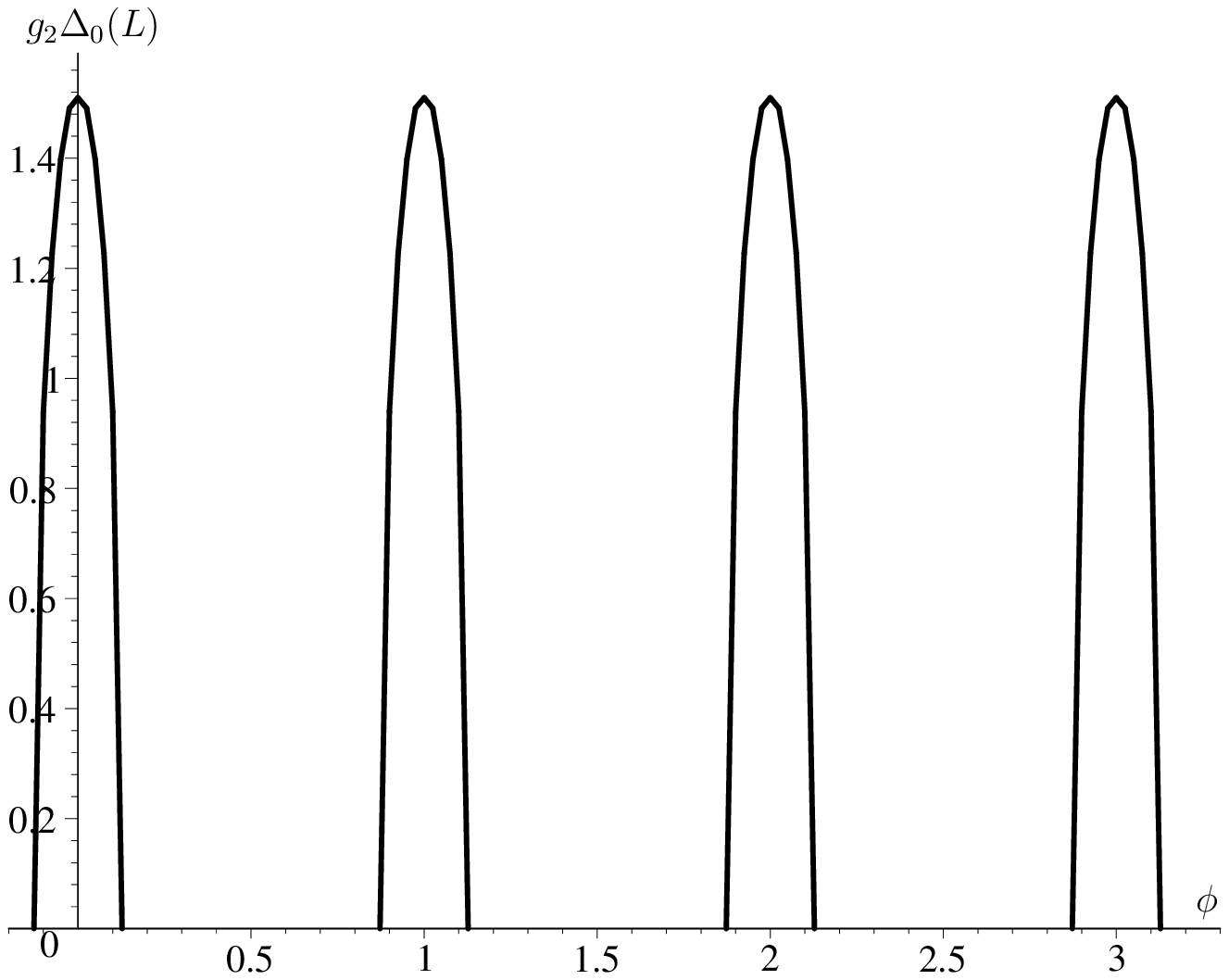}\\
\parbox[t]{0.45\textwidth}{
\caption{The behavior of the gap $M_0(L)$ vs $\phi$ at $g_1=-L/\ln
6$ and arbitrary fixed values of $g_2>0$ or $g_2<g_1$.}   }\hfill
\parbox[t]{0.45\textwidth}{
\caption{The behavior of the gap $\Delta_0(L)$ vs $\phi$ at $g_2=2L$
and arbitrary fixed values of $0<g_1<g_2$.} }
\end{figure}

Finally, we would like to point out another aspect of the  phase
structure of the model (1). It is clear from (\ref{25}) and
(\ref{021}) that at $\mu=0$ the TDP of the system is invariant with
respect to the following simultaneous permutation of coupling
constants and dynamical variables, which is usually called duality
transformation $D$ \cite{thies,ekkz},
\begin{eqnarray}
D:~~g_1\longleftrightarrow g_2,~~ M\longleftrightarrow\Delta.
\label{28}
\end{eqnarray}
Suppose now that at some fixed particular values of the model
parameters, i.e. at $(g_1=A,g_2=B)$, the GMP of the TDP (\ref{021})
lies at the point $(M=M_0,\Delta=\Delta_0)$. Since the TDP is
invariant with respect to duality transformation $D$ (\ref{28}), it is
clear that the permutation of the coupling constant values, i.e. at
$(g_1=B,g_2=A)$, moves the global minimum point of the TDP to the
point $(M=\Delta_0,\Delta=M_0)$. In particular, if at the point
$(g_1=A,g_2=B)$ the superconducting (the chiral symmetry breaking)
phase is realized, then at the point $(g_1=B,g_2=A)$ the chiral
symmetry breaking (the superconducting) phase of the model must be
arranged. As is easily seen from Figs 1-3, just this property of the
phase structure is fulfilled for each figure. Hence, a knowledge of the
phase structure of the model (1) at $g_1<g_2$ is sufficient for
constructing the phase structure at $g_1>g_2$ by taking into account 
the invariance of the TDP under the duality transformation $D$ (\ref{28}.
Thus, there is a duality correspondence between chiral
symmetry breaking and superconductivity in the framework of the
model (1) at $\mu=0$. It is also necessary to remark that in
\cite{thies,ekkz} the CSB-SC duality was established in the framework
of (1+1)-dimensional models with a continuous chiral symmetry group. In
contrast, in the present consideration the duality correspondence is
a property of the model (1) with a discrete $\gamma^5$ chiral symmetry.

\section{Phase structure at $L\ne\infty$ and $\mu\ne 0$}

Numerical investigations show again that a global minimum  point (GMP) of
the TDP (\ref{21}) cannot be located at the point of the form
$(M\ne 0,\Delta\ne 0)$, i.e. at least one of the quantities $M$ and
$\Delta$ is equal to zero in the GMP of the thermodynamic potential
(\ref{21}). So, in order to establish the GMP $(M_0,\Delta_0)$ of this
TDP, it is sufficient to compare the least values
of the simpler functions, ${\cal F}_{1\phi}(M)$ and ${\cal F}_{2\phi}(\Delta)$, which are the reductions of the TDP
$\Omega^{ren}_{L\phi}(M,\Delta)$ (see the relation (\ref{21})) on the
$M$ and $\Delta$ axis, correspondingly. Evidently,
\begin{eqnarray}
{\cal F}_{1\phi}(M)&\equiv&\Omega^{ren}_{L\phi}(M,\Delta=0) =F_{1\phi}(M)-\frac 2L\int\frac{dp_1}{2\pi}\sum_{n=-\infty}^{\infty}\left (\mu-\sqrt{E^2_{nL\phi}+M^2}\right )
\Theta\left (\mu-\sqrt{E^2_{nL\phi}+M^2}\right ),\label{24}\\
{\cal F}_{2\phi}(\Delta)&\equiv&
\Omega^{ren}_{L\phi}(M=0,\Delta)=F_{2\phi}(\Delta)-\frac 1L\int\frac{dp_1}{2\pi}\sum_{n=-\infty}^{\infty}
\Bigg
(\sqrt{(E_{nL\phi}+\mu)^2+\Delta^2}\nonumber\\
&&~~~~~~~~~~~~~~~~~~~~~~~~~~~~~~~~~+\sqrt{(E_{nL\phi}-\mu)^2+\Delta^2}-2\sqrt{E^2_{nL\phi}+\Delta^2}
~\Bigg ), \label{26}
\end{eqnarray}
where $\Theta(x)$ is the Heaviside step function, $F_{1\phi}(M)$ and
$F_{2\phi}(\Delta)$ are presented in (\ref{022}) and (\ref{023}),
respectively, and $E_{nL\phi}=\sqrt{p_1^2+4\pi^2(n+\phi)^2/L^2}.$
Investigating and comparing the behavior of GMPs of the functions
(\ref{24}) and (\ref{26}) vs external parameters $L$, $\mu$, $g_1$,
$g_2$, and $\phi$, it is possible to obtain the phase structure of the
model. Of course, in reality we have studied  numerically the
functions (\ref{24}) and (\ref{26}). The results of our analysis for
typical values of the magnetic flux $\phi$ and chemical potential
$\mu$ (and for arbitrary fixed values of the quantity $L$) are
presented in Figs. 6-8. For example, in Figs. 6 and 7 the
$(g_1,g_2)$-phase structure of the model is presented, respectively,
at $\phi=0$ and $\phi=1/12$. For both figures the chemical potential
values are selected to be the same, i.e. $L\mu=0$, $L\mu=0.2$,
$L\mu=0.4$,  $L\mu=0.6$. In Fig. 8 one can see the $(g_1,g_2)$-phase
portraits of the model at $\phi=1/3$ and for the following set of
the chemical potential values: $L\mu=0$, $L\mu=0.6$, $L\mu=1.2$,
$L\mu=1.8$. In any case, on the basis of these phase portraits it is
easy to see that with growth of the chemical potential $\mu$ (at fixed
$\phi$ and $L$ values) the phase III gradually fills the whole
$(g_1,g_2)$ plane (with the exception of the line $g_2=0$). Namely, 
it is clear from Figs. 6-8 that at an arbitrary fixed
point $(g_1,g_2)$ (note, that $g_2\ne 0$) of a phase diagram there
exists a critical value $\mu_c$ of the chemical potential such that
at $\mu>\mu_c$ the superconducting phase is realized in the system.

In particular, if initially at $\mu=0$ we have a SC ground state,
then $\mu_c=0$. It means that in this case SC is maintained in the
model at arbitrary values of $\mu$. The typical behaviour of the superconducting gap $\Delta_0$ vs $\mu$ in this case is depicted in Fig. 9 for $g_1=-2L$, $g_2=-L$ and at $\phi =0$ (as it is clear from Fig. 6a that for these values of coupling constants we have superconductivity at $\mu=0$).
However, if at $\mu=0$ the point $(g_1,g_2)$ is arranged in the CSB or symmetrical phase, then $\mu_c>0$. The typical behavior of gaps $M_0$ and $\Delta_0$ in this case is
represented by Fig. 10, where a competition between the CSB and SC
order parameters, $M_0$ and $\Delta_0$, is depicted at $g_1=-L$, $g_2=-2L$ and $\phi =0$. It is clear from this figure that there is a critical value $\mu_c\approx 0.49/L$ of the chemical potential, where a first order phase transition occurs from CSB (at $\mu<\mu_c$) to SC phase (at $\mu>\mu_c$).
\begin{figure}
\includegraphics[width=0.45\textwidth]{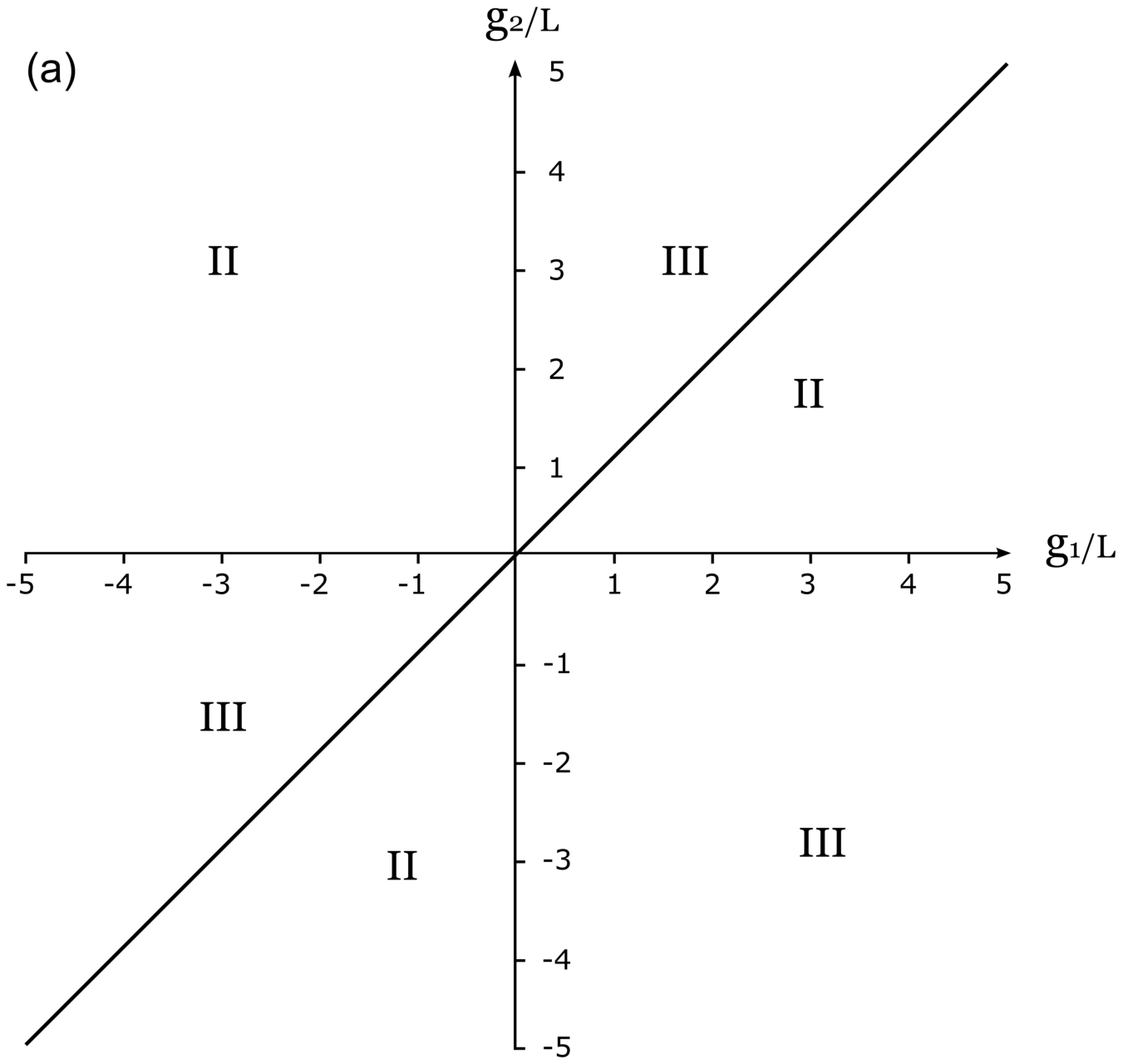}
\hfill
\includegraphics[width=0.45\textwidth]{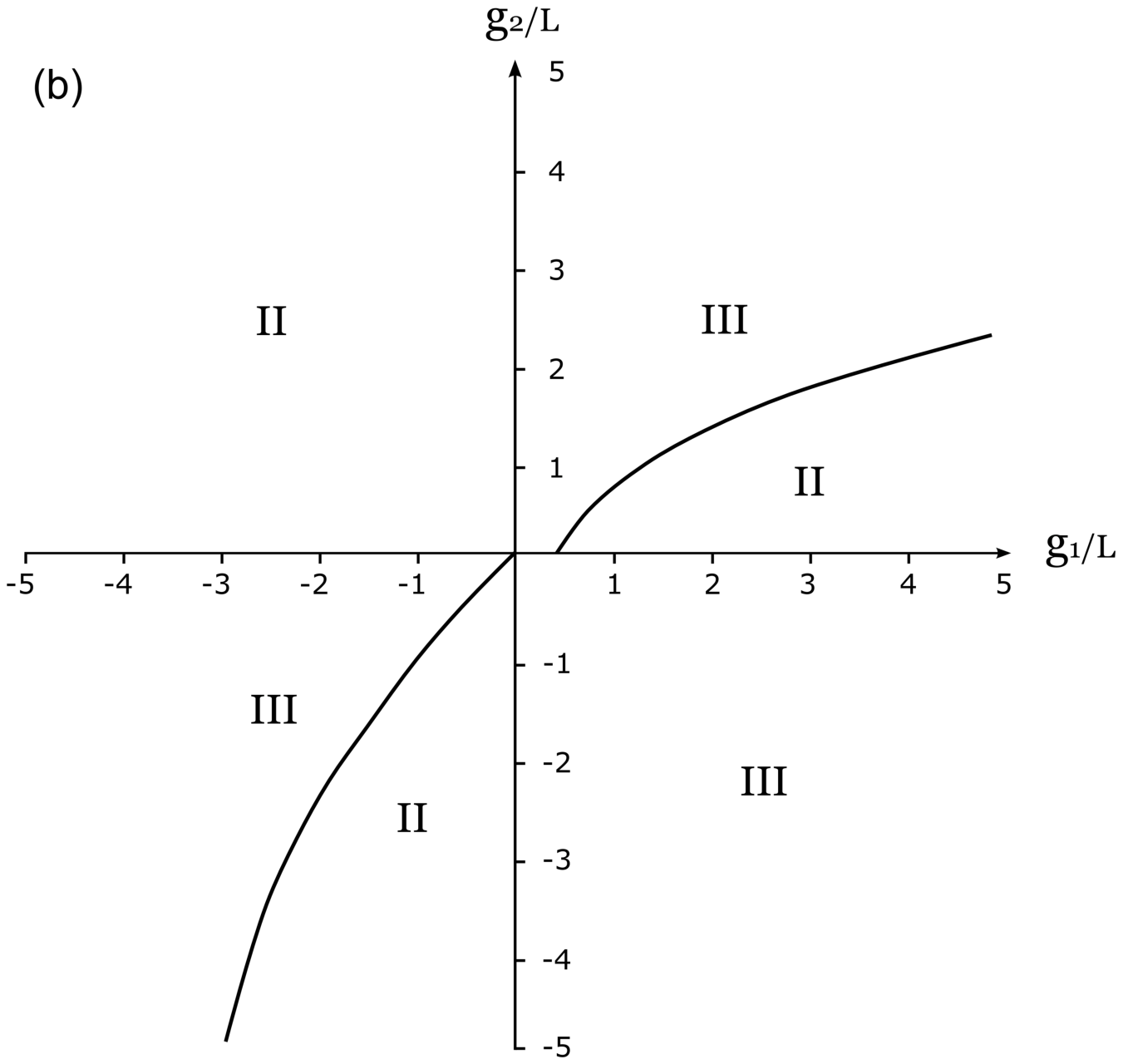}\\
\includegraphics[width=0.45\textwidth]{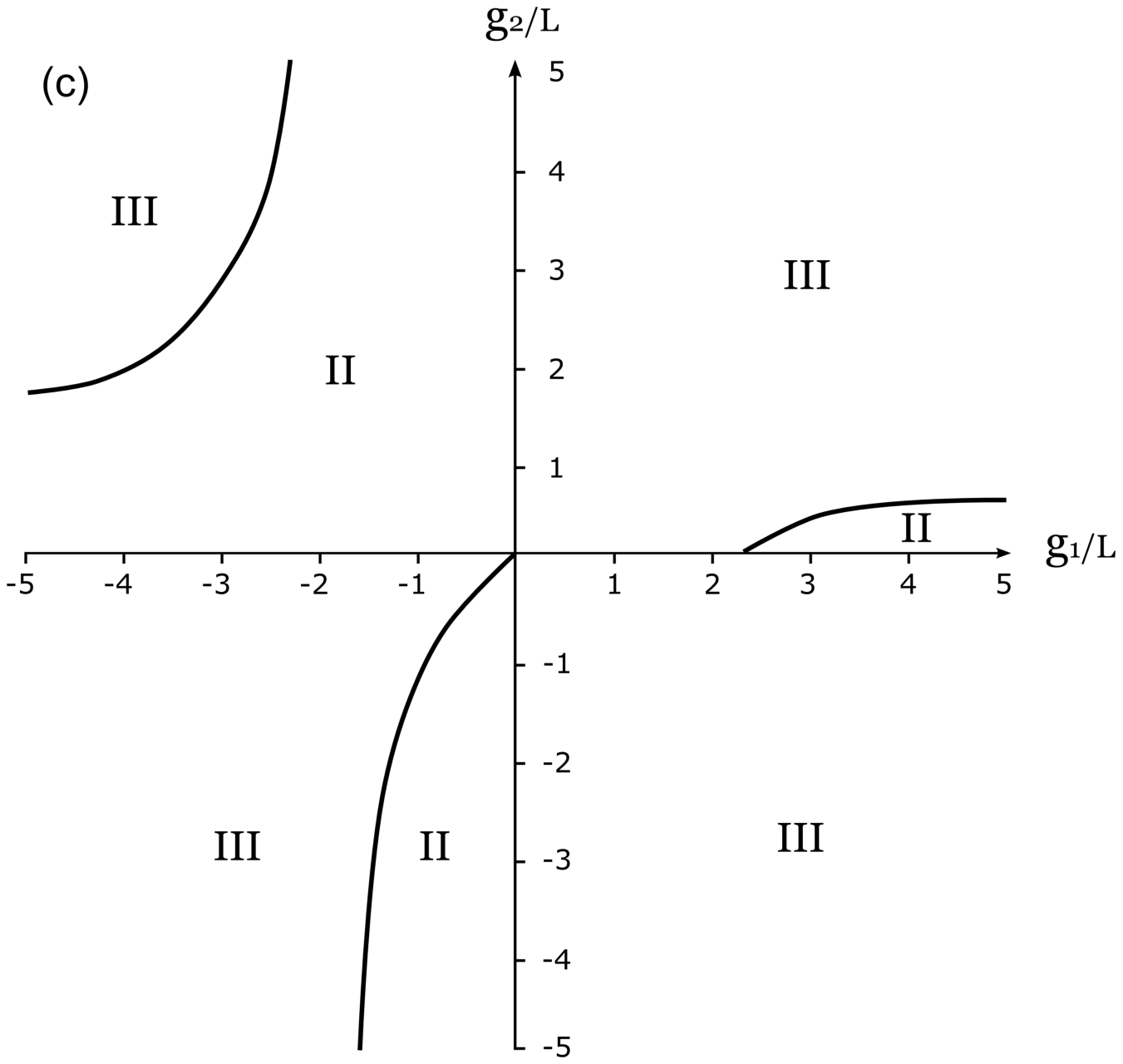}
\hfill
\includegraphics[width=0.45\textwidth]{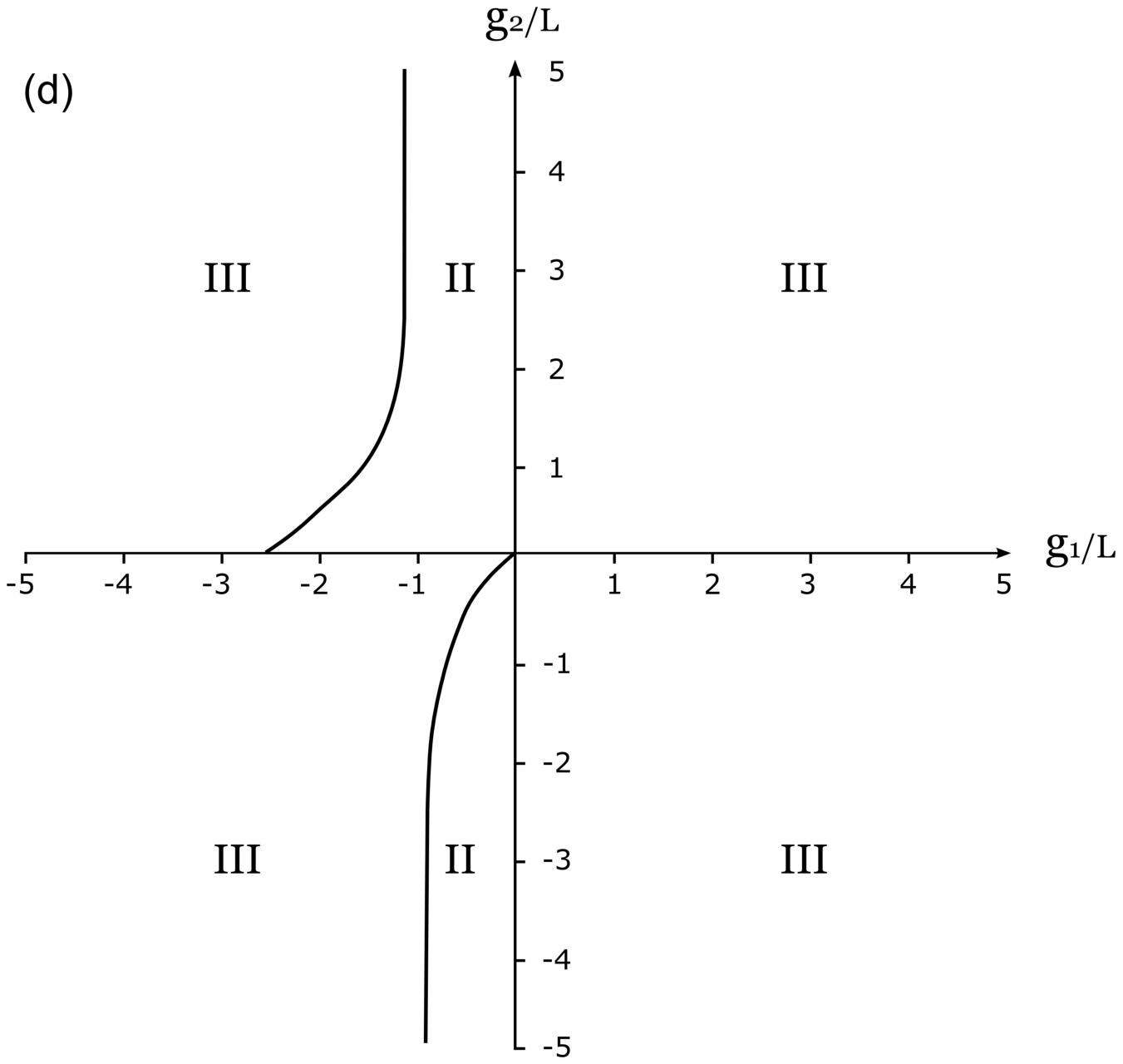}\\
\caption{The $(g_1,g_2)$-phase portrait of the model at $\phi=0$, arbitrary fixed values of $L$ and for different values of chemical potential $\mu$. (a) The case $L\mu=0$. (b) The case  $L\mu=0.2$. (c) The case  $L\mu=0.4$. (d) The case  $L\mu=0.6$. We
use the same designations of the phases as in Fig. 1. }
\end{figure}
\begin{figure}
\includegraphics[width=0.45\textwidth]{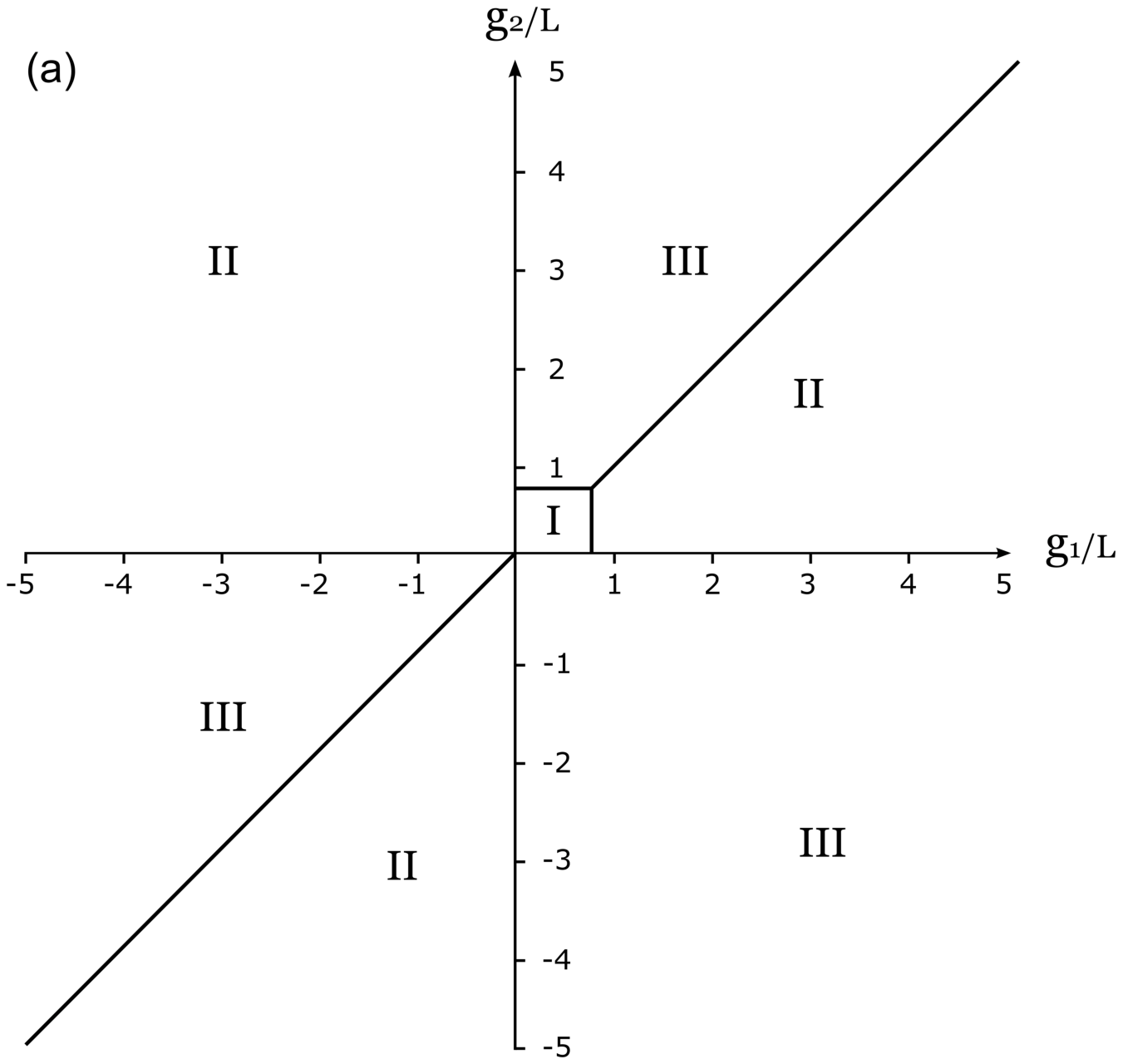}
\hfill
\includegraphics[width=0.45\textwidth]{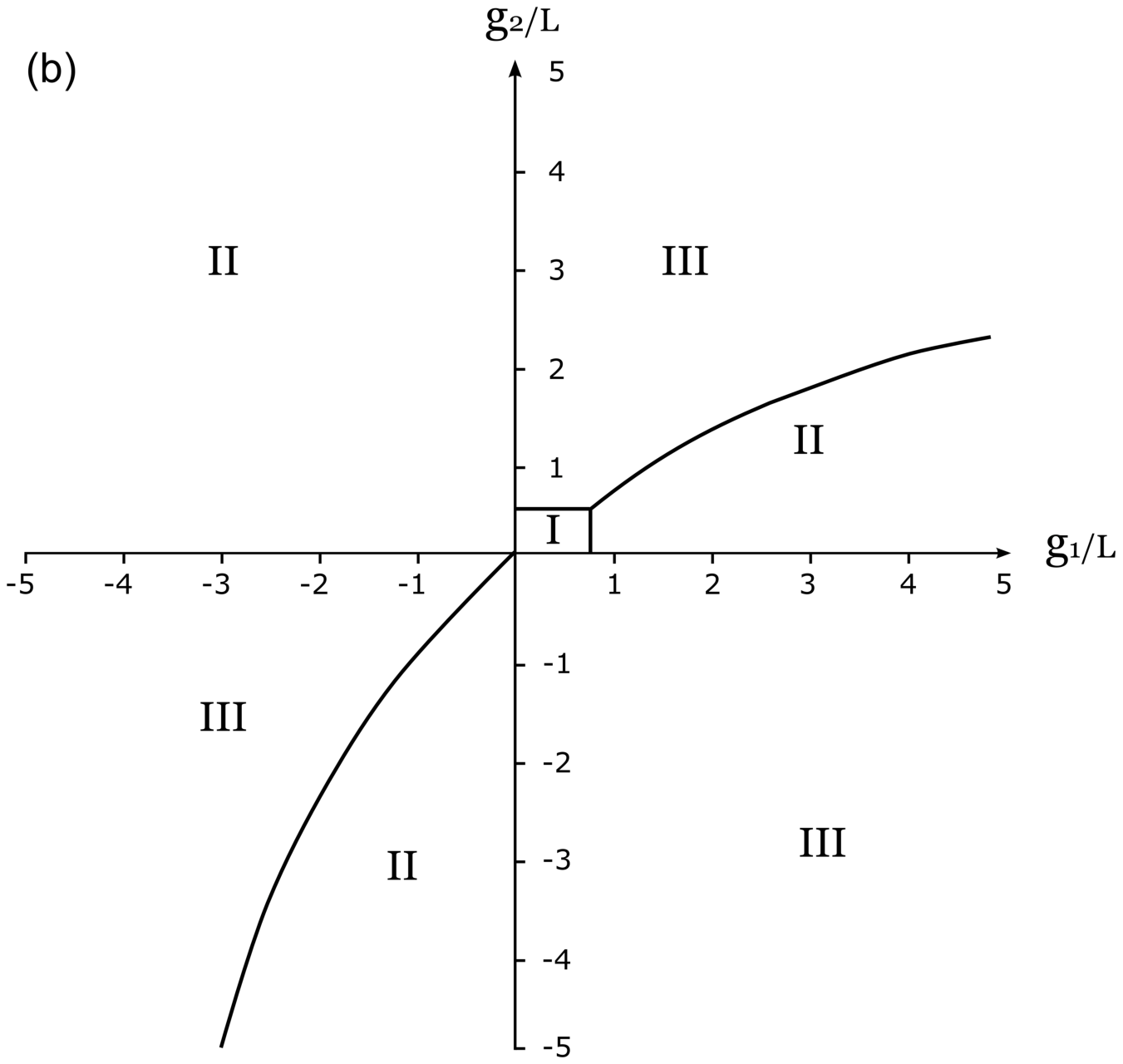}\\
\includegraphics[width=0.45\textwidth]{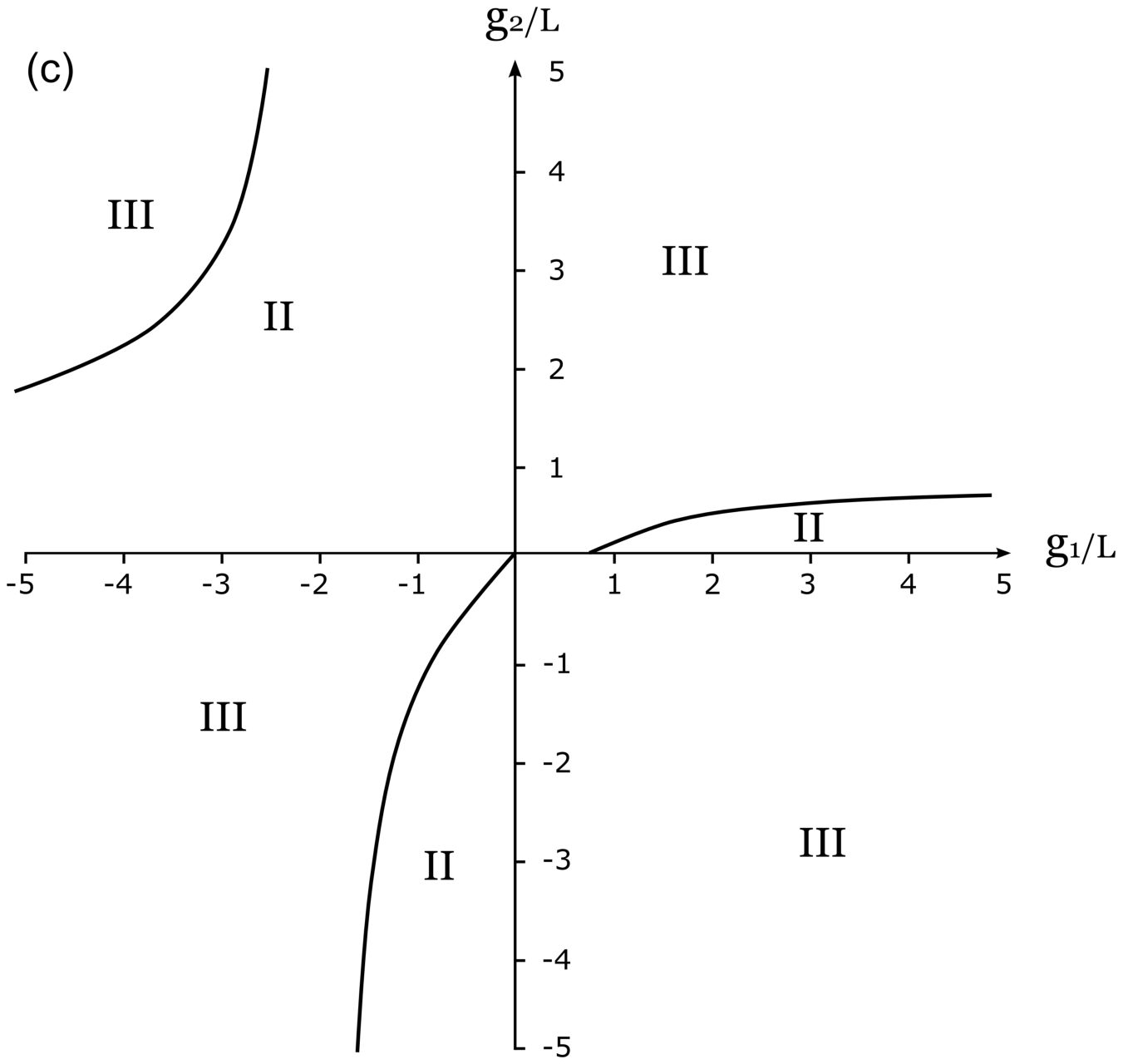}
\hfill
\includegraphics[width=0.45\textwidth]{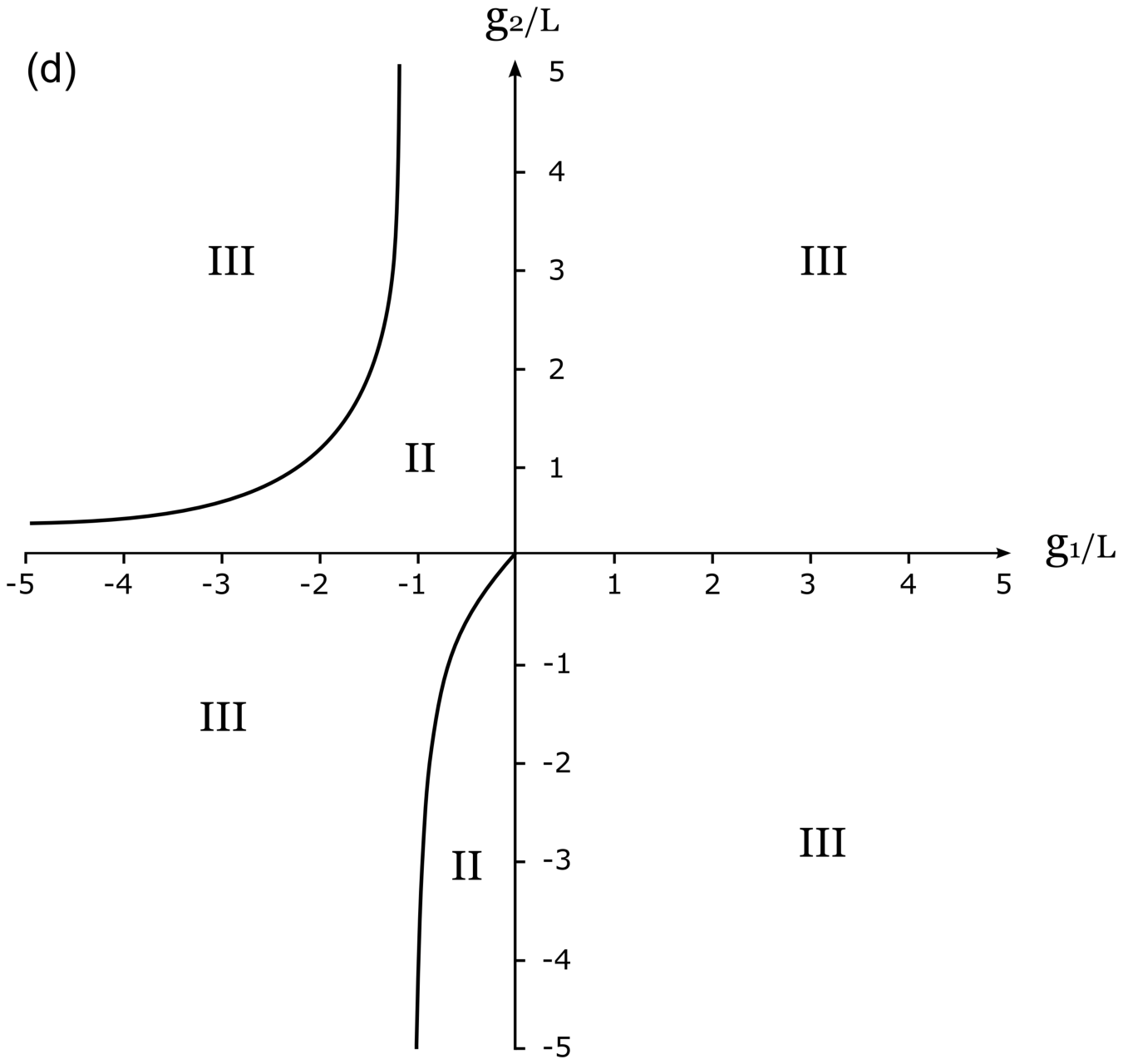}\\
\caption{The $(g_1,g_2)$-phase portrait of the model at $\phi=1/12$, arbitrary fixed values of $L$ and for different values of chemical potential $\mu$. (a) The case $L\mu=0$. (b) The case  $L\mu=0.2$. (c) The case  $L\mu=0.4$. (d) The case  $L\mu=0.6$. We
use the same designations of the phases as in Fig. 1. }
\end{figure}
\begin{figure}
\includegraphics[width=0.45\textwidth]{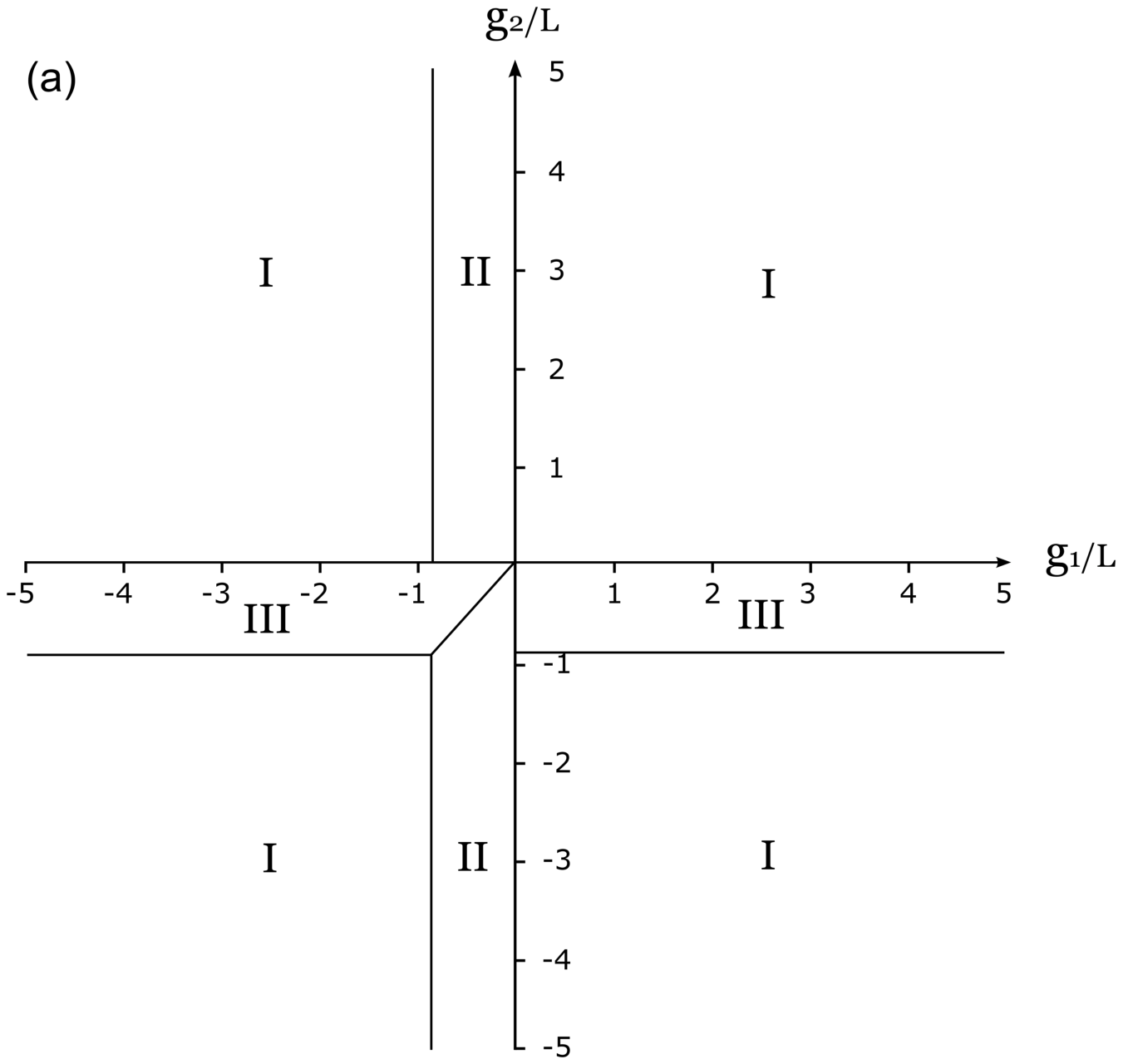}
\hfill
\includegraphics[width=0.45\textwidth]{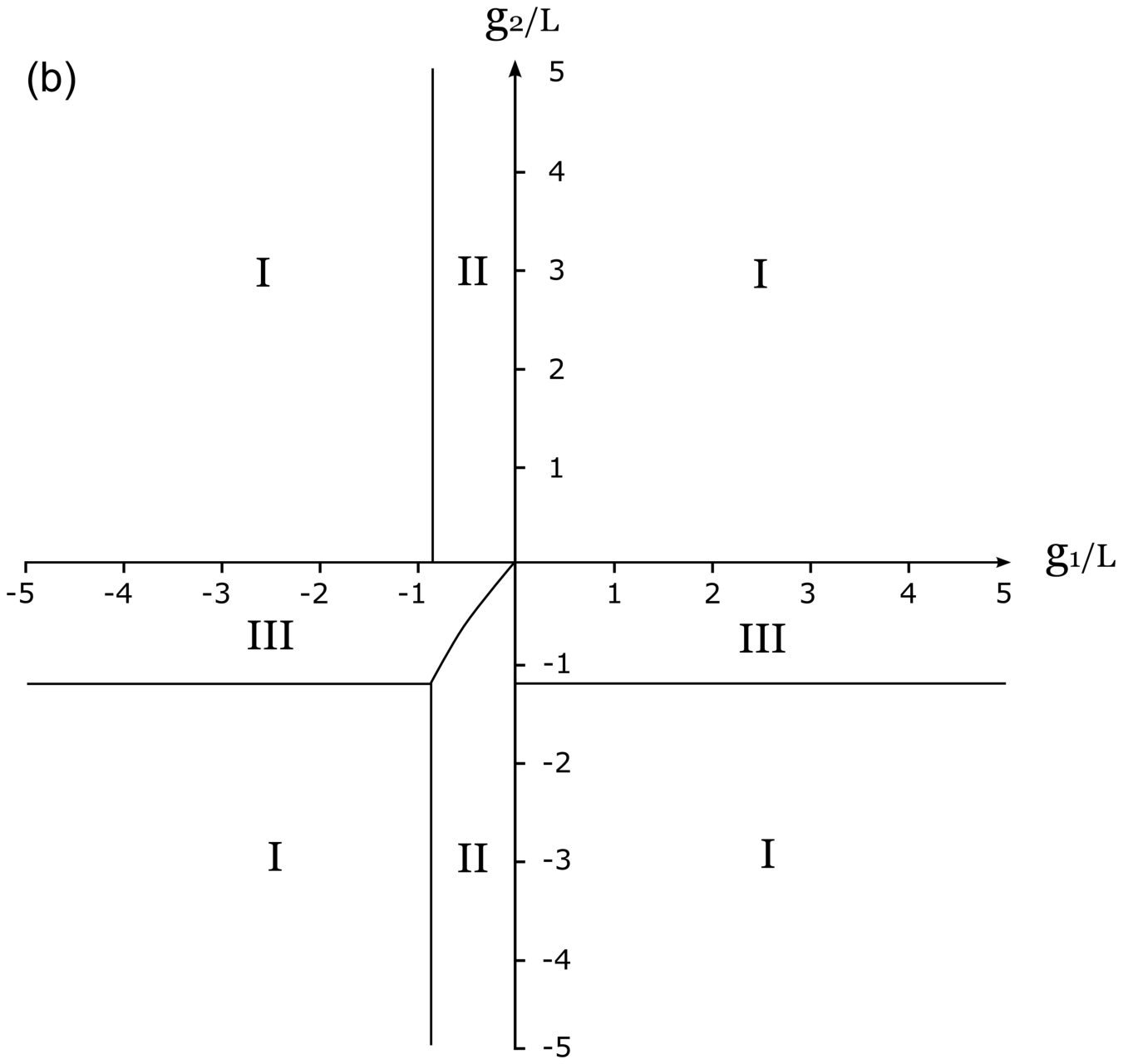}\\
\includegraphics[width=0.45\textwidth]{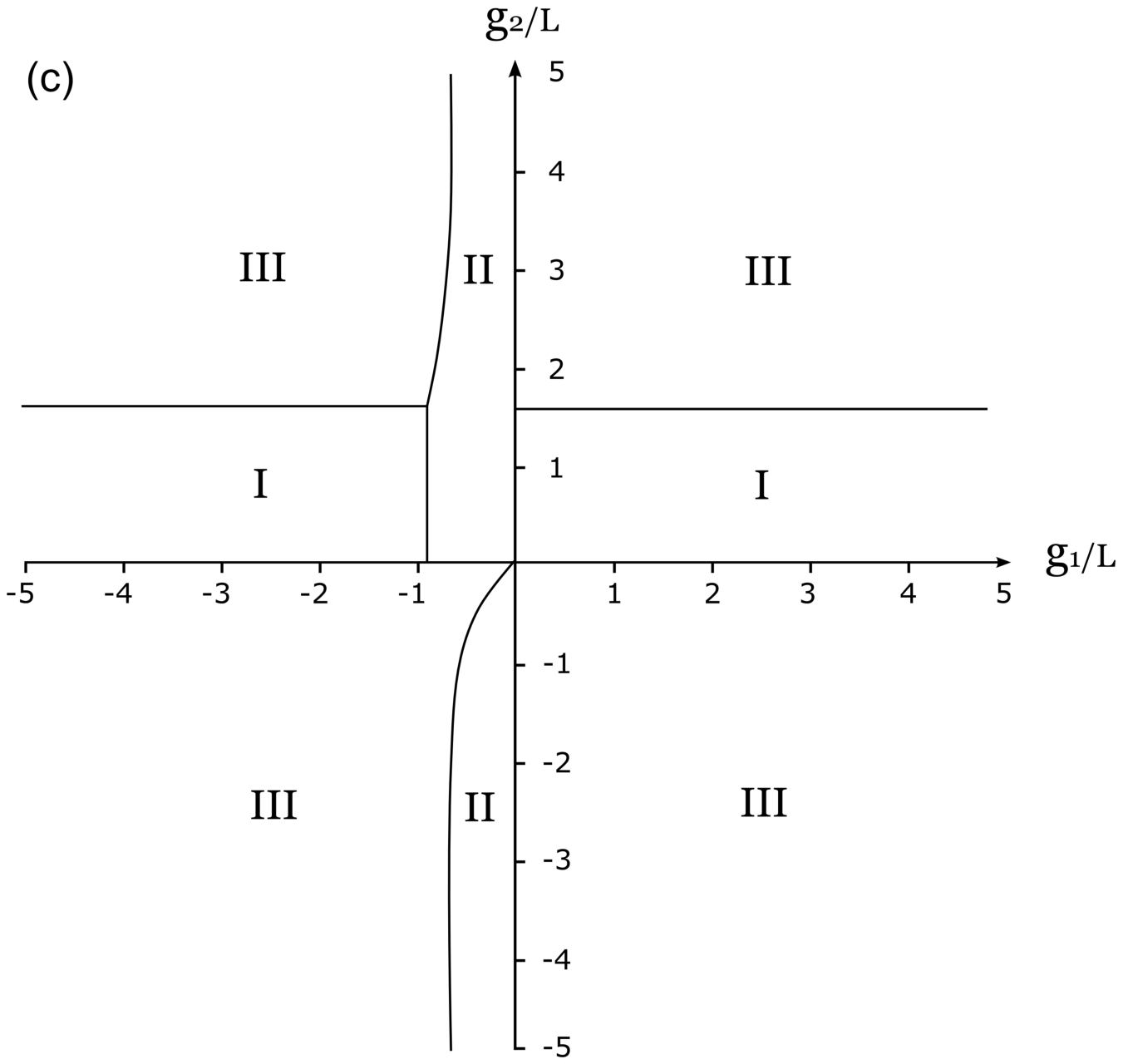}
\hfill
\includegraphics[width=0.45\textwidth]{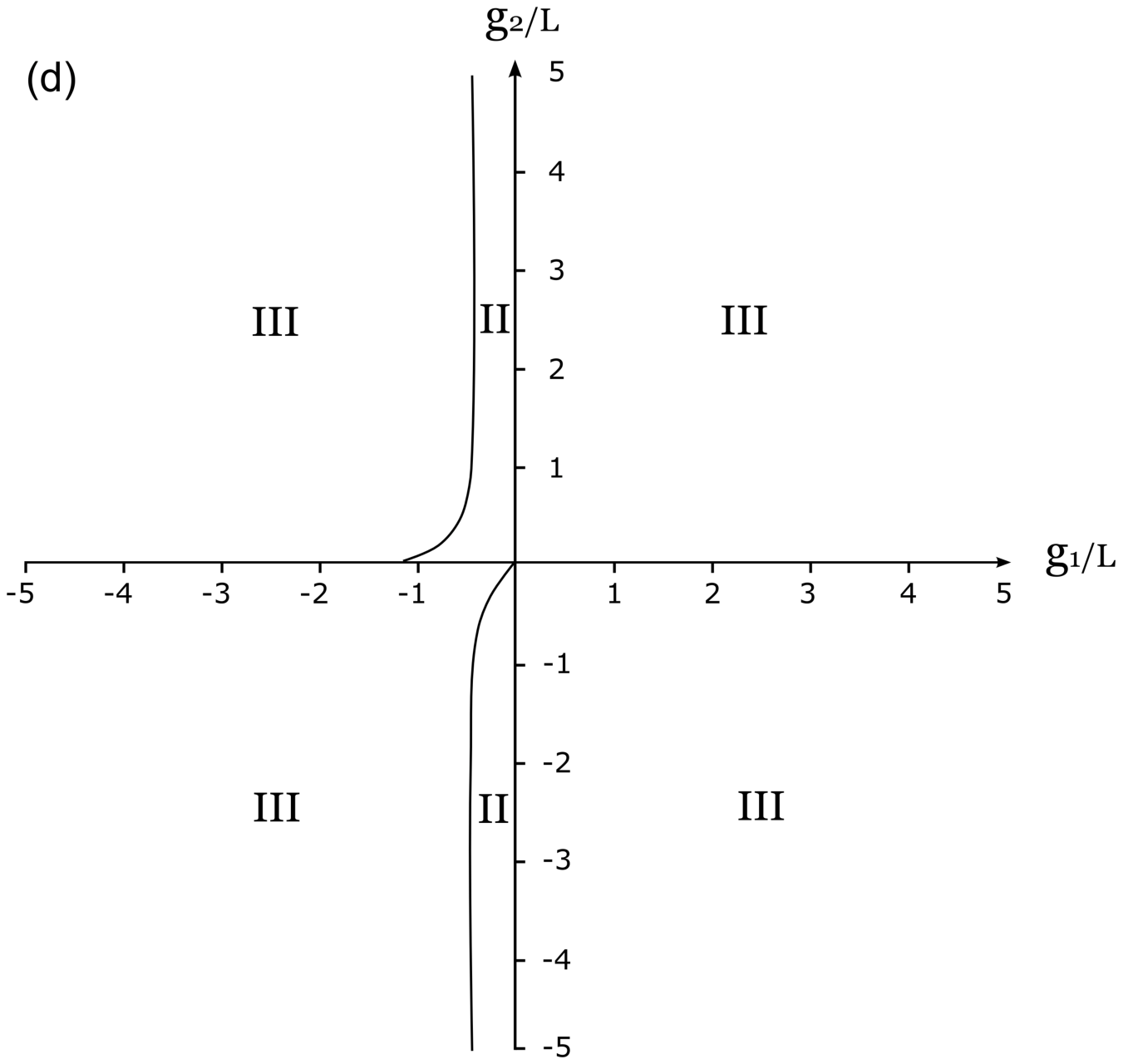}\\
\caption{The $(g_1,g_2)$-phase portrait of the model at $\phi=1/3$, arbitrary fixed values of $L$ and for different values of chemical potential $\mu$. (a) The case $L\mu=0$. (b) The case  $L\mu=0.6$. (c) The case  $L\mu=1.2$. (d) The case  $L\mu=1.8$. We
use the same designations of the phases as in Fig. 1. }
\end{figure}
\begin{figure}
\includegraphics[width=0.45\textwidth]{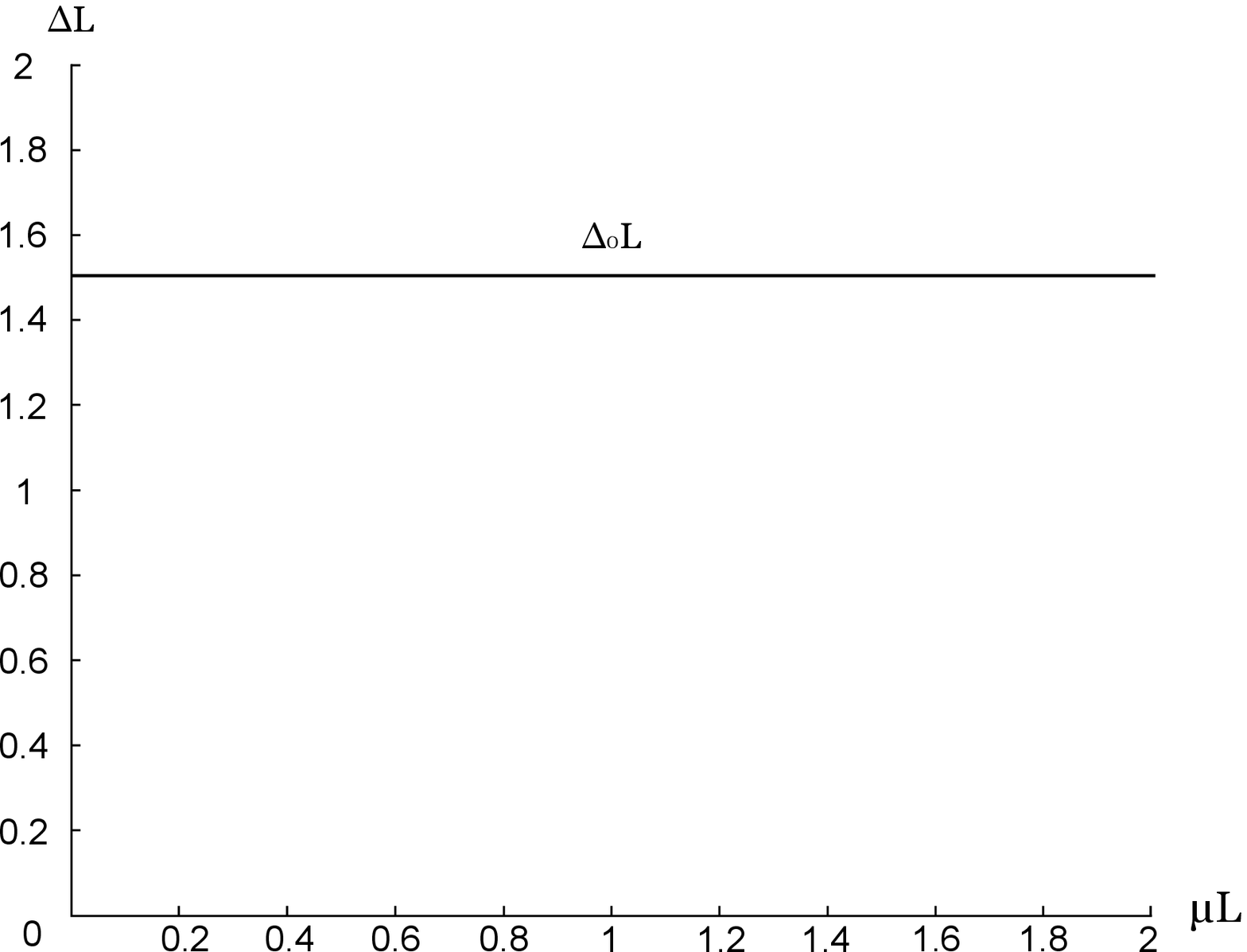}
\hfill
\includegraphics[width=0.45\textwidth]{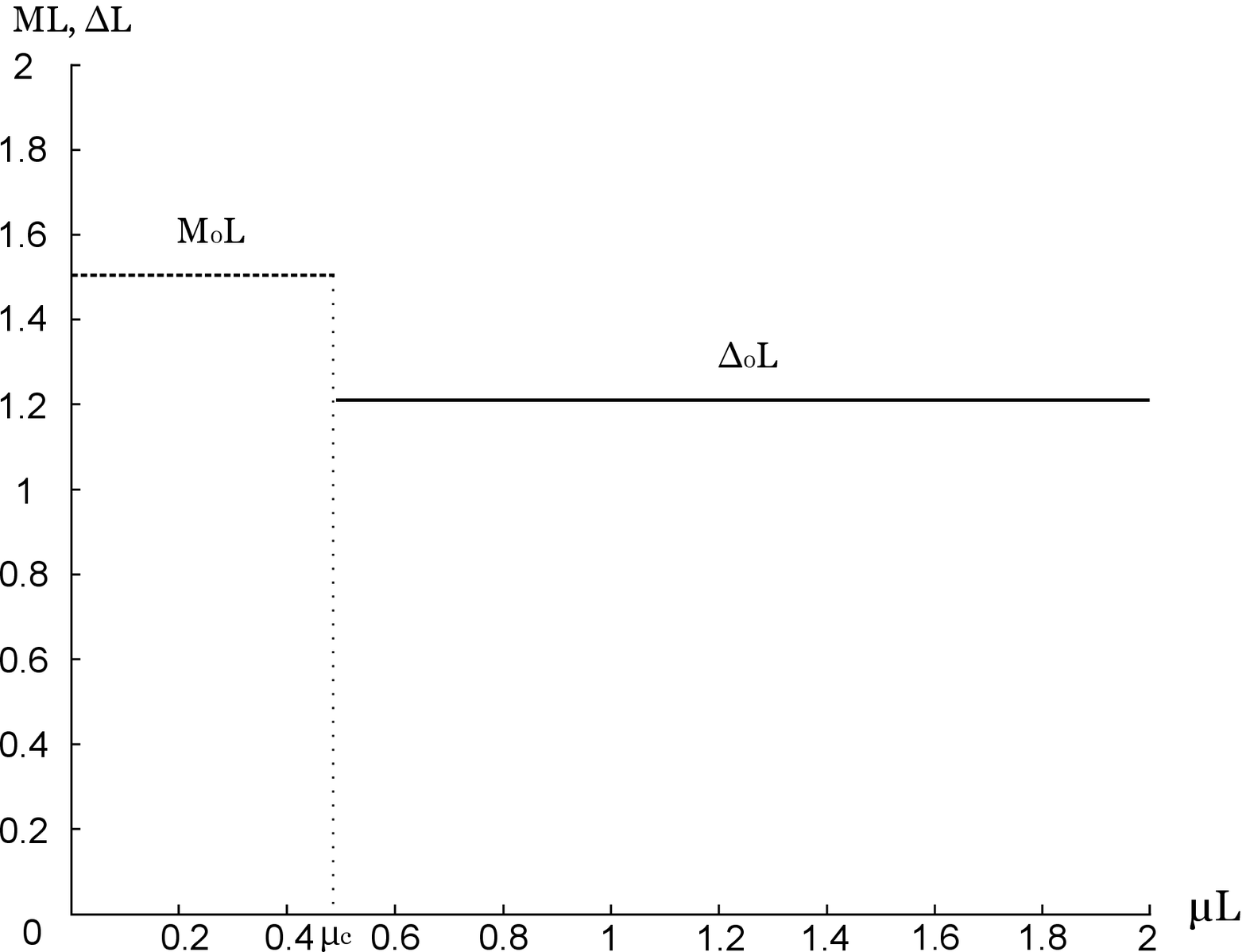}\\
\parbox[t]{0.45\textwidth}{
\caption{The behavior of the gap $\Delta_0$ vs $\mu$ at $g_1=-2L$, $g_2=-L$, $\phi=0$ and arbitrary fixed value of $L$. (In this case $M_0\equiv 0$ vs $\mu$).}   }\hfill
\parbox[t]{0.45\textwidth}{
\caption{The behavior of the gaps $\Delta_0$ and $M_0$ vs $\mu$ at $g_1=-L$, $g_2=-2L$, $\phi=0$ and arbitrary fixed value of $L$. The 1st order phase transition between CSB and SC phases occurs at $\mu_c\approx 0.49/L$.} }
\end{figure}

In the previous Section we have pointed out that for some fixed
points $(g_1,g_2)$ and $\mu=0$ there can appear in the model the
reentering of the CSB or SC phases vs magnetic flux $\phi$. It is
clear from Figs. 6-8 that the reentrance effect takes place for
rather small fixed nonzero values of $\mu$ as well. Indeed, let us
suppose that $L\mu=0.6$ and $(g_1,g_2)$ is fixed in such a way,
e.g., that $g_1/L=g_2/L=-3$. Then at $\phi=0$ and $\phi=1/12$ we
have in this point of the $(g_1,g_2)$-phase diagram the SC phase III
(see Figs. 6d and 7d, correspondingly). However, at $\phi=1/3$ the symmetric phase I
is already realized in this point (see Fig. 8b). Since all physical
quantities of the model are periodical vs $\phi$, one can conclude
that in the above mentioned fixed point $(g_1,g_2)$ and at $L\mu=0.6$ there are both periodical restoration of the initial symmetry and
periodical reentering of the SC phase vs $\phi$. Since at rather
large values of $\mu$ and for arbitrary values of the magnetic 
flux $\phi$ the $(g_1,g_2)$-phase structure of the model looks like the
phase diagram of Fig. 8d with an extremely narrow phase II, it is
necessary to note that the reentrance effect of the model disappears
at sufficiently high values of the chemical potential.

\section{Summary and conclusions}

In this paper we have studied the competition between chiral and
superconducting condensations in the framework of the
(2+1)-dimensional 4FQFT model (1), when one of spatial coordinates
is compactified and the two-dimensional space has $R^1\otimes S^1$
topology (the length of the circumference $S^1$ is $L$). We consider
this $R^1\otimes S^1$ space as a cylinder embedded into the real
flat three-dimensional space. In addition, we supposed that there is
an external magnetic field flux $\Phi$ through a transverse section
of the cylinder (as a result, the boundary conditions (\ref{21}) are
fulfilled, where $\phi=\Phi/\Phi_0$). The model describes interactions both in the fermion-antifermion (or chiral) and superconducting difermion (or Cooper pairing) channels with bare couplings $G_1$ and $G_2$,
respectively. Moreover, it is chirally and $U(1)$ invariant (the
last group corresponds to conservation of the fermion number or
electric charge of the system). To avoid the ban on the spontaneous
breaking of continuous symmetry in (2+1)-dimensional field theories,
we considered the phase structure of our model in the leading
order of the large-$N$ technique, i.e. in the limit $N\to\infty$,
where $N$ is a number of fermion fields, as it was done in the 
(1+1)-dimensional analog of the model \cite{chodos,abreu}. 
The temperature is zero in our consideration.

{\bf The case $L=\infty$, $\mu=0$.} First of all we have
investigated the thermodynamic potential of the model in the flat
two-dimensional space with trivial topology, i.e. at $L=\infty$,
with zero chemical potential $\mu=0$. In this case the phase
portrait is presented in Fig. 1 in terms of the renormalization
group invariant finite coupling constants $g_1$ and $g_2$, defined
in (\ref{018}). Each point $(g_1,g_2)$ of this diagram corresponds
to a definite phase. For example, at $g_{1,2}>0$, i.e. at
sufficiently small values of the bare coupling constants $G_{1,2}$
(see the comment at the end of Section \ref{mu0}), both the discrete
$\gamma^5$ chiral and $U(1)$ symmetries are not violated, and the
system is in the symmetric phase, etc.

{\bf The case $L\ne \infty$, $\mu=0$.} In this case there are two
qualitatively different situations depending on the value of the
magnetic flux $\phi$. Indeed, if $0\le \phi<1/6$, then the typical
$(g_1,g_2)$-phase portrait of the model is presented in Fig. 2, but
if $1/6<\phi<1/2$, then the typical phase portrait of the model is
drawn in Fig. 3. In particular, it follows from Fig. 2 that at
$\phi=0$ (in this case the quantity $g_c$ (\ref{190}) is equal to
zero) we have in the region $g_{1,2}>0$ spontaneous breaking of the
chiral $\gamma^5$ or $U(1)$ symmetry (in contrast, if $L=\infty$
then initial symmetry remains intact in this region). So, the
compactification of the space, i.e. at $L\ne \infty$, induces
spontaneous breaking of the symmetry.

Note also that all physical quantities of the model are periodic
functions vs magnetic flux $\phi$ (see, e.g., Figs. 4 and 5, where
the behavior of the CSB and SC gaps are presented). It is clear from
Fig. 5 that for some points of the $(g_1,g_2)$ plane an increasing
magnetic flux $\phi$ is accompanied by periodical reentrance
of the SC (or CSB) phase. We expect that this effect can be observed
in condensed matter experiments. 
Such a response of physical systems on the action of external magnetic field perpendicular to a direction of compactified coordinate is contrasted with the case, when magnetic field is directed along the compactified coordinate. In the last case the reentrance effect is absent  \cite{45}. 

Finally, it is necessary to note that at finite $L$ and $\mu=0$ there
is a duality between chiral symmetry breaking and superconductivity
(see the relation (\ref{28}). It means that if at the point
$(g_1,g_2)$ of a phase diagram the CSB phase (the SC phase) is
realized, then at the point  $(g_2,g_1)$ one should have the SC phase
(the CSB phase). Just this property of the model is evident from the
phase diagrams of Figs 2 and 3.

{\bf The case $L\ne\infty$, $\mu\ne 0$.} The $(g_2,g_1)$-phase
portraits of the model are presented in this case in Figs 6--8 for the following representative values of the magnetic flux, 
$\phi=0$, $\phi=1/12$ and $\phi=1/3$, correspondingly. Moreover, in each figure four phase diagrams are
drawn for different values of the chemical potential $\mu$. For
example, in Figs. 6 and 7 chemical potential takes such values that
$L\mu=0$, $L\mu=0.2$, $L\mu=0.4$,  and $L\mu=0.6$,
respectively. Comparing at each fixed $\phi$ the phase diagrams
corresponding to different values of $\mu$, it is possible to
establish the following interesting property of the model: at each
fixed point of the $(g_1,g_2)$ plane (such that $g_2\ne 0$) and fixed
$L\ne \infty$ the growth of the chemical potential leads to appearing
of superconductivity in the system. (The same property of the chemical
potential in the framework of the model (1) was established earlier in
our paper \cite{kzz} at $L=\infty$ even at nonzero temperature.) In
particular, it means that if at $\mu=0$ we have in the model a SC
ground state, then at arbitrary values of $\mu>0$ the
superconductivity persists in the system as well. Moreover, if at
$\mu=0$ we have in the model  CSB or symmetrical ground state, then
there is a critical value $\mu_c>0$ of the chemical potential, such
that at $\mu>\mu_c$ initial CSB or symmetrical ground state is
destroyed and the superconductivity appears.
In other words, if in the physical system of fermions described by
Lagrangian (1) and located on a cylindrical surface, there is an
arbitrary small attractive interaction in the fermion-fermion channel,
then it is possible to generate in the system the SC phenomenon by
increasing the chemical potential.

It is necessary to note that the reentrance of the CSB or SC phases vs
$\phi$ is also possible in the model at rather small nonzero values of
$\mu$. However, in this case the reentrance effect disappears at
sufficiently high values of the chemical potential (see the discussion
at the end of Sec. V).

Since the results of the paper are valid for arbitrary values of
$L$, $0<L<\infty$, we hope that our investigations can shed new
light on physical phenomena taking place in nanotubes as well. In
particular, taking into account the remarks made in the footnote
\ref{foot}, it is possible to relate phase diagrams of Figs. 6 and 8
to physical processes in metallic and semiconducting carbon
nanotubes (with zero external magnetic flux), correspondingly.

\appendix

\section{Algebra of the $\gamma$-matrices in the case of SO(2,1) group}
\label{ApA}

The two-dimensional irreducible representation of the 3-dimensional
Lorentz group SO(2,1) is realized by the following $2\times 2$
$\tilde\gamma$-matrices:
\begin{eqnarray}
\tilde\gamma^0=\sigma_3=
\left (\begin{array}{cc}
1 & 0\\
0 &-1
\end{array}\right ),\,\,
\tilde\gamma^1=i\sigma_1=
\left (\begin{array}{cc}
0 & i\\
i &0
\end{array}\right ),\,\,
\tilde\gamma^2=i\sigma_2=
\left (\begin{array}{cc}
0 & 1\\
-1 &0
\end{array}\right ),
\label{A1}
\end{eqnarray}
acting on two-component Dirac spinors. They have the properties:
\begin{eqnarray}
Tr(\tilde\gamma^{\mu}\tilde\gamma^{\nu})=2g^{\mu\nu};~~
[\tilde\gamma^{\mu},\tilde\gamma^{\nu}]=-2i\varepsilon^{\mu\nu\alpha}
\tilde\gamma_{\alpha};~
~\tilde\gamma^{\mu}\tilde\gamma^{\nu}=-i\varepsilon^{\mu\nu\alpha}
\tilde
\gamma_{\alpha}+
g^{\mu\nu},
\label{A2}
\end{eqnarray}
where $g^{\mu\nu}=g_{\mu\nu}=diag(1,-1,-1),
~\tilde\gamma_{\alpha}=g_{\alpha\beta}\tilde\gamma^{\beta},~
\varepsilon^{012}=1$.
There is also the relation:
\begin{eqnarray}
Tr(\tilde\gamma^{\mu}\tilde\gamma^{\nu}\tilde\gamma^{\alpha})=
-2i\varepsilon^{\mu\nu\alpha}.
\label{A3}
\end{eqnarray}
Note that the definition of chiral symmetry is slightly unusual in
three dimensions (spin is here a pseudoscalar rather than a (axial)
vector). The formal reason is simply that there exists no other $2\times 2$ matrix anticommuting with the Dirac matrices $\tilde\gamma^{\nu}$
which would allow the introduction of a $\gamma^5$-matrix in the
irreducible representation. The important concept of 'chiral'
symmetries  and their breakdown by mass terms can nevertheless be
realized also in the framework of (2+1)-dimensional quantum field
theories by considering a four-component reducible representation
for Dirac fields. In this case the Dirac spinors $\psi$ have the
following form:
\begin{eqnarray}
\psi(x)=
\left (\begin{array}{cc}
\tilde\psi_{1}(x)\\
\tilde\psi_{2}(x)
\end{array}\right ),
\label{A4}
\end{eqnarray}
with $\tilde\psi_1,\tilde\psi_2$ being two-component spinors.
In the reducible four-dimensional spinor representation one deals
with (4$\times$4) $\gamma$-matrices:
$\gamma^\mu=diag(\tilde\gamma^\mu,-\tilde\gamma^\mu)$, where
$\tilde\gamma^\mu$ are given in (\ref{A1}). One can easily show, that
($\mu,\nu=0,1,2$):
\begin{eqnarray}
&&Tr(\gamma^\mu\gamma^\nu)=4g^{\mu\nu};~~
\gamma^\mu\gamma^\nu=\sigma^{\mu\nu}+g^{\mu\nu};~~\nonumber\\
&&\sigma^{\mu\nu}=\frac{1}{2}[\gamma^\mu,\gamma^\nu]
=diag(-i\varepsilon^{\mu\nu\alpha}\tilde\gamma_\alpha,
-i\varepsilon^{\mu\nu\alpha}\tilde\gamma_\alpha).
\label{A5}
\end{eqnarray}
In addition to the  Dirac matrices $\gamma^\mu~~(\mu=0,1,2)$ there
exist two other matrices $\gamma^3$, $\gamma^5$ which anticommute
with all $\gamma^\mu~~(\mu=0,1,2)$ and with themselves
\begin{eqnarray}
\gamma^3=
\left (\begin{array}{cc}
0~,& I\\
I~,& 0
\end{array}\right ),\,
\gamma^5=\gamma^0\gamma^1\gamma^2\gamma^3=
i\left (\begin{array}{cc}
0~,& -I\\
I~,& 0
\end{array}\right ),\,\,
\label{A6}
\end{eqnarray}
with  $I$ being the unit $2\times 2$ matrix.

\section{Proper-time representation of the TDP (\ref{14}) }
\label{ApC}

Let us derive another expression for the unrenormalized TDP $V^{un}(M,\Delta)$ which is  equivalent to (\ref{14}), by using the Schwinger
proper-time method. Here and in the next appendix we use the
general relation
\begin{eqnarray}
\sqrt{A}=\frac 1{\sqrt{\pi}}\int_0^\infty\frac{ds}{s^2}\Big
(1-e^{-s^2A}\Big ),
\label{C1}
\end{eqnarray}
where $A>0$ and an improper integral in the right-hand side is obviously a
convergent one. Supposing that $A=A_\pm\equiv\sqrt{|\vec p|^2+(M\pm\Delta)^2}$, one can use the relation (\ref{C1}) in (\ref{14}) and find
\begin{eqnarray}
V^{un}(M,\Delta)=
\frac{M^2}{4G_1}+\frac{\Delta^2}{4G_2}&+&
\frac 1{\sqrt{\pi}}\int\frac{d^2p}{(2\pi)^2}\Bigg (\sum_\pm \int_0^\infty \frac{ds}{s^2}e^{- s^2\left
[p_1^2+p_2^2+(M\pm\Delta)^2\right ]}~\Bigg ),
\label{C2}
\end{eqnarray}
where we have omitted an unessential infinite constant, which does
not depend on dynamical variables $M$ and $\Delta$. (Due to this
reason the proper-time integral in (\ref{C2}) and below, in
(\ref{C3}), is divergent.) Integrating in (\ref{C2}) over $p_1$ and
$p_2$, we obtain finally the following proper-time expression for
the unrenormalized TDP (\ref{14}):
\begin{eqnarray}
V^{un}(M,\Delta)=
\frac{M^2}{4G_1}+\frac{\Delta^2}{4G_2}&+&
\frac 1{4\pi^{3/2}}\sum_\pm \int_0^\infty \frac{ds}{s^4}e^{- s^2
(M\pm\Delta)^2}.
\label{C3}
\end{eqnarray}

\section{Derivation of expression (\ref{B8}) for the
thermodynamic potential}
\label{ApB}

Let us denote by $A_{n,\pm}$ the following expression:
\begin{eqnarray}
A_{n,\pm}=p_1^2+\frac{4\pi^2(n+\phi)^2}{L^2}+(M\pm\Delta)^2.
\label{B1}
\end{eqnarray}
Then the TDP (\ref{20}) has the form:
\begin{eqnarray}
V^{un}_{L\phi} (M,\Delta)= \frac{M^2}{4G_1}+\frac{\Delta^2}{4G_2}&-&
\frac 1L\int\frac{dp_1}{2\pi}\sum_{n=-\infty}^{\infty}\Bigg
(\sum_\pm\sqrt{A_{n,\pm}}~\Bigg ). \label{B2}
\end{eqnarray}
To proceed, it is very convenient to use for square roots
$\sqrt{A_{n,\pm}}$ in (\ref{B2}) the proper-time representation
(\ref{C1}). Then, up to an infinite constant independent on
dynamical variables $M$ and $\Delta$, we have
\begin{eqnarray}
V^{un}_{L\phi} (M,\Delta)= \frac{M^2}{4G_1}+\frac{\Delta^2}{4G_2}&+&
\frac
1{L\sqrt{\pi}}\sum_\pm\int\frac{dp_1}{2\pi}\sum_{n=-\infty}^{\infty}
\int_0^\infty \frac{ds}{s^2}e^{- s^2\left
[p_1^2+(M\pm\Delta)^2+\frac{4\pi^2}{L^2}(n+\phi)^2\right ]}.
\label{B3}
\end{eqnarray}
First of all let us sum in (\ref{B3}) over $n$, taking  into account
there the well-known Poisson summation formula,
\begin{eqnarray}
\sum_{n=-\infty}^{\infty} e^{- s^2\frac{4\pi^2}{L^2}(n+\phi)^2}=
\frac{L}{2\pi}\frac{\sqrt{\pi}}{s}\sum_{n=-\infty}^{\infty}
e^{-\frac{n^2L^2}{4s^2}}e^{i2\pi n\phi}=
\frac{L}{2\pi}\frac{\sqrt{\pi}}{s}\biggl\{1+2\sum_{n=1}^{\infty}
e^{-\frac{n^2L^2}{4s^2}}\cos(2\pi n\phi)\biggr\}. \label{B4}
\end{eqnarray}
Then, after integration in the obtained expression over $p_1$, we have
\begin{eqnarray}
V^{un}_{L\phi} (M,\Delta)=V^{un}(M,\Delta) +\frac
1{2\pi^{3/2}}\sum_{\pm}\sum_{n=1}^{\infty}\int_0^\infty
\frac{ds}{s^4}e^{- s^2(M\pm\Delta)^2-\frac{n^2L^2}{4s^2}}\cos(2\pi
n\phi), \label{B5}
\end{eqnarray}
where $V^{un}(M,\Delta)$ is the proper-time represented  effective
potential of the model in the vacuum (\ref{C3}). Taking into account
the relation
\begin{eqnarray*}
\int\limits^{\infty}_{0}
dxx^{\nu -1} e^{-\frac{a}{x}-b x}=
2\left (\frac ab \right )^{\nu /2}K_\nu(2\sqrt{ab}),
\end{eqnarray*}
it is possible to integrate in (\ref{B5}) over $s$,
\begin{eqnarray}
V^{un}_{L\phi} (M,\Delta)=V^{un}(M,\Delta) +\frac
1{2\pi^{3/2}}\sum_{\pm}\sum_{n=1}^{\infty}\left
(\frac{2|M\pm\Delta|}{nL}\right )^{3/2}K_{-\frac 32}\Big
(nL|M\pm\Delta|\Big )\cos(2\pi n\phi), \label{B6}
\end{eqnarray}
where $K_\nu(z)$ is the third order modified Bessel function, and
\begin{eqnarray}
K_{-\frac 32}(z)=K_{\frac 32}(z)=-\sqrt{\frac{\pi z}{2}}\frac d{dz}\left (\frac{e^{-z}}{z}\right )=\sqrt{\frac{\pi z}{2}}e^{-z}\frac{z+1}{z^2}.\label{B7}
\end{eqnarray}
Using the relation (\ref{B7}) in (\ref{B6}),  we obtain the
expression (\ref{B8}) for the unrenormalized effective potential at
$L\ne\infty$ and $\mu=0$.

\end{document}